\documentclass[aps,prd,preprint]{revtex4}
\usepackage{multirow}
\usepackage{graphicx,psfrag}
\usepackage{amsmath}
\usepackage{bm}
\usepackage{float}
\usepackage{hyperref}

\newcommand{\drawsquare}[2]{\hbox{%
\rule{#2pt}{#1pt}\hskip-#2pt					
\rule{#1pt}{#2pt}\hskip-#1pt					
\rule[#1pt]{#1pt}{#2pt}}\rule[#1pt]{#2pt}{#2pt}\hskip-#2pt	
\rule{#2pt}{#1pt}}						

\begin{document}

\vspace*{2cm}

\title{\Large Towers of baryons of the \mbox{${\cal N}=2$} quark model band in \\ the large $N_c$ limit}

\author{C.T. Willemyns$^{\,a,b}$, N.N. Scoccola$^{\,a,b,c}$}
\affiliation{$^{a}$ CONICET, Rivadavia 1917, (1033) Buenos Aires, Argentina}
\affiliation{$^{b}$ Physics Department, Comisi\'{o}n Nacional de Energ\'{\i}a At\'{o}mica,\\Av.\ Libertador 8250, (1429) Buenos Aires, Argentina}
\affiliation{$^{c}$ Universidad Favaloro, Sol\'{\i}s 453, 1078 Buenos Aires, Argentina}

\begin{abstract} 

We performed a complete analysis of the spectrum of all the states in the \mbox{${\cal N}=2$} quark model harmonic oscillator band with \mbox{$N_f=3$} in the \mbox{large $N_c$} limit including the often disregarded antisymmetric multiplet $[{\bf 20},1^+]$. We included configuration mixing effects.
We found that the states in the $[{\bf 56},L^+]$ and $[{\bf 70},L^+]$ with $L=0,2$ fall into nine towers of degenerate states.
We found that nonstrange antisymmetric states fall into three towers and respond to the same structure as the states in the $[{\bf 70},1^-]$ multiplet.
We also showed explicitly the compatibility of these results and the scattering resonance picture.

\end{abstract}


\maketitle

\section{Introduction}

Baryon spectroscopy has been essential for our understanding of QCD in the low-energy, strong-coupling regime.
In this context, the quark model for baryons has since long time been a useful tool to analyze the spectrum and properties of excited baryons \cite{Capstick:2000qj}. In the quark model, baryon resonances belong to \mbox{$SU(2N_f)\times O(3)$} representations that are accommodated into ${\cal N}$ bands of the harmonic oscillator.
Recent studies of lattice QCD calculations \cite{Edwards:2011jj,Edwards:2012fx} seem to confirm this classification scheme, strongly suggesting a connexion between QCD and the quark model.
The main advantage of these numerical lattice calculations is that they rely entirely on the fundamental QCD theory.
On the other hand, the lattice QCD method lacks the transparency and simplicity of an analytic approach.
To unravel a physical
picture in terms of effective degrees of freedom, effective
interactions, and symmetries is still a challenging task.
Thus, despite the continuing improvements in the lattice QCD techniques, the understanding of resonant state properties from first principles remains a very hard problem.

The \mbox{large $N_c$ QCD} approach suggested by 't Hooft \cite{tHooft:1973alw} has become a powerful tool to understand the spectrum and properties of ground state baryons and their first excited states.
This approach is based on the result that in the sector of the ground state light flavored baryons, there is a contracted $SU(2N_f)_c$ spin-flavor symmetry in  the limit of \mbox{large $N_c$} \cite{Dashen:1993ac,Dashen:1993as}.

Since the \mbox{large $N_c$} picture was first used to describe baryons by Witten \cite{Witten:1979kh}, the $1/N_c$ expansion using effective quark operators has been  applied  with  great  success  to describe properties of the  ground  state baryons (see Ref.~\cite{Manohar:2002sx} for a brief review and references therein).
The ground state baryons  belong to the ${\cal N}=0$  band of the quark model classification scheme and are described  by  the  symmetric representation $\bf 56$ of $SU(6)$ for $N_f=3$.
Excited states require a more complex treatment as they also appear in mixed-symmetric and anti-symmetric representations of
$SU(2N_f)$. Several detailed studies of the masses of excited baryons in the ${\cal N}=1$ band, which belong to the $[{\bf  70},1^-]$ multiplet with  mixed spin-flavor symmetry, have been carried out with great success using a mass operator built with core and excited quark operators \cite{Pirjol:2003ye,Goity:2002pu,Schat:2001xr,Carlson:1998vx,Pirjol:1997bp,Pirjol:1997sr,Goity:1996hk}.
Multiplets belonging to the ${\cal N} =2$ band have also been separately studied, namely the $[{\bf 56},0^+]$ multiplet in Ref.~\cite{Carlson:2000zr}, the $[{\bf 56},2^+]$ multiplet in Ref.~\cite{Goity:2003ab} and the baryons of the $[{\bf  70},L^+]$ with $L=0,2$ in Refs.~\cite{Matagne:2005gd,Matagne:2006zf}.
In addition to the analysis of the mass spectra, strong and electromagnetic decays were also studied in the $1/N_c$ expansion approach (see Ref.~\cite{Matagne:2014lla} for a recent review).

As already mentioned, the classification scheme for baryon resonances based on irreducible representations of \mbox{$SU(2N_f)\times O(3)$} originates from the quark model. However, physical states appear as combinations of these quark model irreducible representations, this fact is usually known as configuration mixing. The \mbox{$SU(2N_f)\times O(3)$} symmetry is not something that follows from the fundamental QCD theory.
This is also manifest in \mbox{large $N_c$ QCD} where the configuration mixing effects are not $N_c$ suppressed \cite{Goity:2004pw,Goity:2005fj,Cohen:2003fv}.
Instead of  what is predicted by the quark model, states in the \mbox{large $N_c$} limit belong to irreducible representations of a contracted $SU(2N_f)_c$ symmetry and organize into towers labeled by the associated quantum number $K$.
Despite not being a suppressed effect, the works in Refs.~\cite{Carlson:2000zr,Goity:2003ab,Matagne:2005gd,Matagne:2006zf} do not include configuration mixing in the \mbox{large $N_c$} limit. Only recently such an effect was included in an study of the ${\cal N}=2$ band nonstrange states ~\cite{Willemyns:2017jnh}. However, the antisymmetric multiplet was not included in that work. In this context,
the analysis for $N_f=3$ of the complete set of multiplets of the ${\cal N}=2$ band within the $1/N_c$ expansion
and including the configuration mixing effects appears to be relevant. This is, in fact, the main  aim of the present work.

The tower structure is well understood in the nonstrange case, the $K$ number that arises from the \mbox{large $N_c$} limit is the spin vector $\bf K$ which is known as ``grand spin'' in chiral soliton models. In the nonstrange case, ${\bf K}={\bf J}+{\bf I}$ where $J$ is the total spin and $I$ is the isospin. The $K$ number holds a very simple relation with the orbital angular momentum number $L$: ${\bf K}={\bf L}$ for the symmetric representations and ${\bf K}={\bf L}+{\bf 1}$ for the mixed-symmetric ones \cite{Willemyns:2017jnh}. However, for $N_f=3$ the content of each tower cannot be found so easily since the relation of $K$ to the states quantum numbers is more complicated.

In addition to the quark operator method mentioned above, there is another natural approach to excited baryons from a \mbox{large $N_c$} perspective known as the resonance picture \cite{Cohen:2005yj}.
While ground-state baryons are stable in the \mbox{large $N_c$} limit, excited baryons are all resonances.
To analyze baryon resonances it is relevant to study scattering processes,
such as meson-nucleon scattering, in channels for which
such resonances may reveal themselves.
It is important to point out that, from a $N_c$ counting point of view, the resonance width of baryons scales as $N_c^0$ \cite{Witten:1979kh}, so that the existence of well-defined narrow baryon states is not ensured at \mbox{large $N_c$}; however, we can rely on the fact that the empirical evidence indicates detectable resonances.
The resonance picture is derived entirely from \mbox{large $N_c$ QCD} and contains information on the tower classification. This picture has proven to be a fruitful method to obtain insight into aspects of baryon resonances in a systematic and model independent way \cite{Cohen:2003tb,Cohen:2005ct,Cohen:2003nn}.

In this paper, we study the complete spectrum of the ${\cal N}=2$ band which contains five multiplets: $[{\bf 56},0^+]$, $[{\bf  70},0^+]$, $[{\bf 56},2^+]$, $[{\bf  70},2^+]$, and $[{\bf 20},1^+]$.
We first consider states belonging to the $[{\bf 56},L^+]$ and $[{\bf  70},L^+]$ multiplets with $L=0,2$ extending the work of Ref.~\cite{Willemyns:2017jnh} to $N_f=3$ flavors. We include configuration mixing by generalizing the effective quark operators to mix the $SU(6)\times O(3)$ multiplets to leading order.
The multiplet $[{\bf 20}, 1^+]$ is considered separately since, as it will become clear in Sec.\ref{Sec7}, no configuration mixing between this multiplet and the others is observed within the present scheme.
Baryon states of the antisymmetric representation are often dismissed based on a lack of evidence.
However, states which might be identified with $N_{1/2}$  have been detected in \mbox{$\pi$ + $N$} scatterings and $J/\psi$ decay processes have shown some evidence of the detection of nucleons $N_{3/2}$ associated to the antisymmetric representation.
These states correspond to the three star $N(2100)1/2+$ and the one star $N(2040)3/2+$ listed in Ref.~\cite{Tanabashi:2018}.
For all the ${\cal N}=2$ band states we also analyze the mass spectra with the resonant approach and check the compatibility between the two pictures.

This paper is organized as follows. In Sec.~\ref{Sec2}, we describe the building of the symmetric and mixed-symmetric baryon states; in Sec.~\ref{Sec3}, we present the effective mass operator and in
 Sec.~\ref{Sec4} we discuss the mass matrices found for $[{\bf 56},L^+]$ and $[{\bf  70},L^+]$ states and we present the spectrum obtained.
In Sec.~\ref{Sec5} we discuss the tower structure found for the $[{\bf 56},L^+]$ and $[{\bf  70},L^+]$ states in the context of large $N_c$ QCD.
In Sec.~\ref{Sec6} we describe the method used to analyze excited baryons with a meson scattering approach, we present our results and associate them with our operator analysis results.
Section ~\ref{Sec7} presents the analysis of the $[{\bf 20},1^+]$ multiplet.
In Sec.~\ref{Sec8} we summarize our conclusions.
App.~\ref{AppC} provides details of the calculations performed to obtain the effective operators matrix elements.
In App.~\ref{AppB} we list the partial-wave amplitudes containing the resonances of the $[{\bf 56},L^+]$ and $[{\bf  70},L^+]$ multiplets along with the \mbox{large $N_c$} mass eigenvalues found in the $1/N_c$ expansion.
In App.~\ref{AppE} we present details of the calculations to obtain the core composition of the antisymmetric states.
In App.~\ref{AppB2} we list the partial-wave amplitudes containing the resonances of the antisymmetric states along with the \mbox{large $N_c$} mass eigenvalues found for the nonstrange states in the $1/N_c$ expansion.
In App.~\ref{AppD} we give the explicit expressions of the reduced matrix elements for mixed-symmetric core states.

\section{Symmetric and mixed-symmetric states}\label{Sec2}

States of the ${\cal N}=2$ band can be analyzed as three quark systems with $N_c=3$. For $N_f=3$ these states belong to irreducible representations of $ SU(6) \otimes O(3)$ where $SU(6)$ contains the flavor group $SU(3)$ and the spin group $SU(2)$.
For $N_c=3$ only three spin-flavor representations occur: completely symmetric (S), mixed symmetric (MS) and completely antisymmetric (A).
Each symmetry corresponds to the $SU(6)$ multiplets  $\bf 56$, ${\bf 70}$ and ${\bf 20}$ respectively.
In this section, we consider the S and MS representations, the building of states in the antisymmetric representation is described in Sec.~\ref{Sec7}.

The analysis of spin-flavor multiplets in \mbox{large $N_c$} is a straightforward extension of methods familiar from $N_c=3$.
In the $N_c>3$ generalization one assumes that the additional $N_c-3$ quarks appear in a completely symmetric spin-flavor combination.
This generalization of the quark model has the same emergent symmetries as \mbox{\mbox{large $N_c$} QCD}. Thus it is an efficient way to deduce group-theoretical results.

In the \mbox{large $N_c$} approach, the multiplets have an infinite number of baryons, the physical baryons can be identified with states at the top of the flavor representations while the other states are spurious baryons that are not relevant when $N_c=3$. In addition to these spurious states that appear in the multiplets containing physical states, when \mbox{$N_c>3$} additional spin-flavor representations arise, these multiplets contain only spurious states that also decouple in the physical limit $N_c=3$. However, spurious states should be considered when $N_c$ is arbitrary since with this approach states associated with physical baryons can result in a combination that contains spurious states, as long as their quantum numbers allow it.
Relevant \mbox{large $N_c$} spin-flavor multiplets are those having states with the same quantum numbers as those associated with the physical states, this means that we have to consider all multiplets containing states with a (total) spin $J$, a hypercharge $Y$ and an isospin  number $I$ that correspond to a baryon observed in Nature.
To identify these states, we must analyze the decomposition of $SU(6)$ spin-flavor representations into separate $SU(2)$ spin and $SU(3)$ flavor representations for $N_c$ quark baryons.
Relevant $SU(3)$ flavor representations that emerge from the \mbox{large $N_c$} generalization are
\begin{eqnarray}
 ``{\bf 8}"\equiv \left(1,\frac{N_c-1}{2}\right),
 ``{\bf 10}"\equiv \left(3,\frac{N_c-3}{2}\right),
 ``{\bf 1}"\equiv \left(0,\frac{N_c-3}{2}\right),
 ``{\bf S}"\equiv \left(2,\frac{N_c-5}{2}\right),
\end{eqnarray}
where we used the $SU(N_f)$ Dynkin label which consists of a multiplet $(n_1, n_2,..., n_{N_f-1})$ where the non-negative integers $n_r$ stand for the number of boxes in row $r$ of the Young diagram that exceed the number of boxes in row $r+1$.
The labels chosen to dub flavor and spin-flavor representations are given by the dimension of the representation when \mbox{$N_c=3$}; hence the quotation marks.
The representation $``{\bf S}"$ emerges only when $N_c>3$ but contains states that could potentially mix with those associated with physical states if the flavor symmetry were broken, namely $\Sigma^{\bf S}$, $\Xi^{\bf S}$ and $\Omega^{\bf S}$. The supra-index indicate the flavor representation, we shall omit quotation marks in these cases to lighten notation.
The $``{\bf 1}"$ representation is a singlet for $N_c=3$ but not for arbitrary $N_c$, in particular, for $N_c=5$ this irreducible representation is $(0,1)\equiv\bar{\bf 3}$, this means that it also contains spurious states in addition to  $\Lambda^{\bf1}$ associated with the physical state, in particular, $\Xi^{\bf1}$ has the quantum numbers $I,Y$ of physical states. Three extra irreducible representations that have a counterpart in $N_c=3$ also arise, $``{\bf 27}"\equiv \left(2,\frac{N_c+1}{2}\right)$, $``{\bf \bar{10}}"\equiv \left(0,\frac{N_c+3}{2}\right)$ and $``{\bf 35}"\equiv \left(4,\frac{N_c-1}{2}\right)$, but as will be clear in the following, they are not contained in the $SU(6)$ multiplets of interest.

The SU(6) decompositions relevant to this work for each spin-flavor multiplet can be found using the general method of Ref.~\cite{Hagen:1965} and are given by
\begin{eqnarray}\label{states2}
\begin{array}{lcl}
 \left[``{\bf56}",0\right]:& & \left[1/2 ,``{\bf8}"\right]^{1/2} \oplus \left[3/2,``{\bf10}"\right]^{3/2} \oplus \dots
  \\
 \left[``{\bf56}",2\right]:& & \left[3/2,``{\bf8}"\right]^{1/2} \oplus \left[5/2,``{\bf8}"\right]^{3/2} \oplus \left[1/2,``{\bf10}"\right]^{3/2} \oplus \left[3/2,``{\bf10}"\right]^{3/2} \oplus \\
 & & \left[5/2,``{\bf10}"\right]^{3/2} \oplus \left[7/2,``{\bf10}"\right]^{3/2} \oplus \dots
  \\
 \left[``{\bf70}",0\right]:& & \left[1/2,``{\bf8}"\right]^{1/2} \oplus \left[3/2,``{\bf8}"\right]^{3/2} \oplus \left[1/2,``{\bf10}"\right]^{1/2} \oplus \left[1/2,``{\bf1}"\right]^{1/2} \oplus  \\
 & & \left[1/2,``{\bf S}"\right]^{1/2*} \oplus \left[3/2,``{\bf S}"\right]^{3/2*} \oplus \left[3/2,``{\bf 10}"\right]^{3/2*} \oplus \left[5/2,``{\bf 10}"\right]^{5/2*} \oplus \dots
  \\
 \left[``{\bf70}",2\right]:& & \left[3/2,``{\bf8}"\right]^{1/2} \oplus \left[5/2,``{\bf8}"\right]^{1/2} \oplus \left[1/2,``{\bf8}"\right]^{3/2} \oplus \left[3/2,``{\bf8}"\right]^{3/2} \oplus \\
 & & \left[5/2,``{\bf 8}"\right]^{3/2} \oplus \left[7/2,``{\bf 8}"\right]^{3/2} \oplus \left[3/2,``{\bf 10}"\right]^{1/2} \oplus \left[5/2,``{\bf 10}"\right]^{1/2} \oplus \\
 & & \left[3/2,``{\bf1}"\right]^{1/2} \oplus \left[5/2,``{\bf1}"\right]^{1/2} \oplus \left[3/2,``{\bf S}"\right]^{1/2*} \oplus \left[5/2,``{\bf S}"\right]^{1/2*} \oplus \\
 & & \left[1/2,``{\bf S}"\right]^{3/2*} \oplus \left[3/2,``{\bf S}"\right]^{3/2*} \oplus \left[5/2,``{\bf S}"\right]^{3/2*} \oplus\left[7/2,``{\bf S}"\right]^{3/2*} \oplus \\
 & & \left[1/2,``{\bf 10}"\right]^{3/2*} \oplus \left[3/2,``{\bf 10}"\right]^{3/2*} \oplus \left[5/2,``{\bf 10}"\right]^{3/2*} \oplus \left[7/2,``{\bf 10}"\right]^{3/2*} \oplus \\
 & & \left[1/2,``{\bf 10}"\right]^{5/2*} \oplus \left[3/2,``{\bf 10}"\right]^{5/2*} \oplus \left[5/2,``{\bf 10}"\right]^{5/2*} \oplus \left[7/2,``{\bf 10}"\right]^{5/2*} \oplus \dots
 \end{array}
\end{eqnarray}
where the notation adopted is $ \left[J,{\bf R}\right]^S$ with ${\bf R}$ being the flavor representation, $J$ the total spin given by the vector sum ${\bf J}={\bf S}+{\bf L}$ and $S$ the spin of the multiplet.
A complete list of the representations contained in the $``{\bf 56}"$ and $``{\bf 70}"$ multiplets can be found in a general form in Eqs.~(3.1) and (3.2) of Ref.~\cite{Cohen:2003nn} for arbitrary $N_f$. Irreducible representations marked with a * contain only spurious states, note that these are not only $``{\bf S}"$ multiplets but also $``{\bf8}"$ and $``{\bf10}"$ with high spin which also decouple in the physical limit \cite{Cohen:2006en}. We only show irreducible representations that contain at least one state with the same $Y,I,J$ quantum numbers as a state of $N_c$ quarks associated with a physical baryon; hence the ellipses. So we have 40 spin-flavor multiplets containing 146 isospin degenerate states to be considered.

Baryons are assigned to states belonging to $ SU(6) \otimes O(3)$, then their wave functions can be expressed with the use of Clebsch-Gordan coefficients as
\begin{equation}
 |(L,S)J,J_z;{\bf R},Y,I,I_z\rangle = \sum_{L_z,S_z}
 \left(
    \begin{array}{cc|c}
        S   	& L 	& J   \\
        S_z 	& L_z    & J_z
    \end{array}
\right)
 |S,S_z;{\bf R},Y,I,I_z\rangle
 |L,L_z \rangle.
\end{equation}

In this approach, as explained in detail in Ref.~\cite{Carlson:1998vx}, excited baryons with $N_c$ quarks of the S and MS representations are composed of a symmetric core of $N_c-1$ quarks and an orbitally excited quark.
Note, however, that the spin-flavor wave functions with defined core spin $S_c$ do not have definite symmetry.
Every baryon state with definite spin-flavor symmetry is a linear combination of states with different core spin $S_c$ that satisfies the relation ${\bf S}={\bf S}_c+{\bf 1/2}$. Then, we have
\begin{equation}\label{states4}
 |S,S_z;(p,q),Y,I,I_z\rangle=\sum_{\eta =\pm 1/2} c_{sym}(p,S,\eta)|S,S_z;(p,q),Y,I,I_z;S_c=S+\eta\rangle,
\end{equation}
where $(p,q)$ stands for representation $\bf R$ in Dynkin notation. The coefficients $c_{sym}(p,S,\eta)$ depend on the $SU(6)$ symmetry of the considered baryon.
Since the core is in an $SU(6)$ symmetric irreducible representation with $N_c-1$ quarks, the Dynkin weights corresponding to the flavor symmetry of the cores are completely determined by $S_c$ and are given by \mbox{$(p_c,q_c)=\left(2S_c,\frac{N_c-1}{2}-S_c\right)$}.
The coefficients giving the correct linear combination to build symmetric and mixed-symmetric states in Eq.~(\ref{states4}) are given by \cite{Goity:2002pu}
\begin{eqnarray}\label{states5}
 c_{MS}(p,S,\pm1/2)&=& \left\{
    \begin{array}{cl}
      1		& \text{if } p=2S\pm1 \text{ or } p=2S\pm 2 , \\
      0		& \text{if } p=2S\mp1 \text{ or } p=2S\mp 2 , \\
      \pm\sqrt{\frac{(2S+1\mp1)(N_c+1\pm(2S+1))}{2N_c(2S+1)}}	& \text{if } p=2S ,  \\
    \end{array}\right.\\
 c_{S}(p,S,\pm1/2)&=&\sqrt{\frac{(2S+1\pm1)(N_c+1\mp(2S+1))}{2N_c(2S+1)}}.\nonumber
\end{eqnarray}
This implies that only MS multiplets with $p\neq2S$ have core states with well-defined spin.

\section{Mass operators}\label{Sec3}

We build the mass operators as described in Ref.~\cite{Carlson:1998vx} but considering a generalization similar to the one used for decay processes and used in Ref.~\cite{Willemyns:2017jnh}.
This generalization consists in considering a generic spatial operator $\xi$ (instead of the orbital excitation operator $\ell$) which admits $L\to L'$ transitions. This allows to include the leading order effects of configuration mixing of spin-flavor representations with different orbital angular momentum.

The building blocks for the construction of the effective mass operators are the \mbox{$ SU(6) \otimes O(3)$} generators: the generic spatial operator of rank $k$ that we denote $\xi^{(k)}$ and the $S^{[1,{\bf 1}]}_i$, $T^{[0,{\bf 8}]}_a$, $G^{[1,{\bf 8}]}_{ia}$ operators associated to the spin-flavor symmetry group. The supra-index in brackets over the spin-flavor operators indicate how they transform in the spin and flavor spaces respectively. The $SU(2N_f)$ Lie algebra commutation relations are given by
\begin{eqnarray}
 [S_i,T_a]&=&0 \,, \nonumber \\
 {}[S_i,S_j]&=& i \epsilon_{ijk} S_k,\,\,\,\,\, [T_a,T_b]=i f_{abc}\ T_c\,,\\
 {}[S_i,G_{ja}]&=& i \epsilon_{ijk} G_{ka},\,\,\,\,\, [T_a,G_{ib}]=i f_{abc}\ G_{ic}\,,\nonumber \\
 {}[G_{ia},G_{jb}]&=&\frac i 4 \delta_{ij}\ f_{abc}\ T_c+\frac{i}{2N_f}\delta_{ab}\ \epsilon_{ijk}\ S_k+\frac i 2 \epsilon_{ijk}\ d_{abc}\ G_{kc}\,.\nonumber
\end{eqnarray}
As explained in the previous section, in the present scheme \mbox{large $N_c$} baryons arise from a generalization in which the states have a symmetric core coupled to an excited quark.
Then, one can define separate one-body operators that act on the core $(S_c)_i$, $(T_c)_a$, $(G_c)_{ia}$ and operators denoted with lower case $s_i$, $t_a$, $g_{ia}$ that act on the excited quark. Since the cores are symmetric, core operators satisfy the operator reduction rules for the ground state \cite{Dashen:1994qi} by replacing $N_c$ by $N_c-1$.

The Hamiltonian can be expressed as a linear combination of effective operators up to order $O(N_c^0)$:
\begin{equation}\label{operators1}
 H=\sum_{i=1}^{5} c_i^{\bf T,T'}O_i + {\cal O}(1/N_c) \ ,
\end{equation}
where ${\bf T,T'}$ stand for the $SU(6)\times O(3)$ irreducible representations $\left[``{\bf56}",L^+\right]$ and $\left[``{\bf70}",L^+\right]$
which we indicate as $S_L$ and $MS_L$ respectively to lighten notation.
Each $O_i$ operator in Eq.~(\ref{operators1}) is constructed using the building blocks mentioned above and considering the reduction rules  for the core operators. There are five spin-singlet flavor-singlet operators that contribute to the baryon masses up to order ${O}(N_c^0)$ given by
\begin{eqnarray}\label{operators2}
 O_1 &=& \left(N_c \openone\right)^{[0,{\bf1}]},	\nonumber\\
 O_2 &=& \left(\xi^{(1)} s\right)^{[0,{\bf1}]}, \nonumber\\
 O_3 &=& \frac{1}{N_c}\left(\xi^{(2)}\left(g G_c\right)^{[2,{\bf 1}]}\right)^{[0,{\bf1}]}, \\
 O_4 &=& \frac{1}{N_c}\left(\xi^{(1)}\left(t G_c\right)^{[1,{\bf 1}]}\right)^{[0,{\bf1}]},\nonumber\\
 O_5 &=& \frac{1}{N_c} \left(t T_c\right)^{[0,{\bf1}]}.\nonumber
\end{eqnarray}
Operators on Eq.~(\ref{operators2}) are a generalization of operators on Ref.~\cite{Cohen:2005ct} in a slightly modified basis as we favoured a simpler form of the expressions for $O_4$ and $O_5$ instead of substracting the contributions to the nonstrange matrix elements.

\section{Mass matrices}\label{Sec4}

In this section, we present  the calculations of the mass matrix elements of Eq.~(\ref{operators1}) for the states in the multiplets given in Eq.~(\ref{states2}).

The ${\cal O}_i$ operators of Eq.~(\ref{operators2}) can all be written in a general form as $\left(\xi^{(l)} {\cal G}^{[s,{\bf r}]}\right)^{[j,{\bf r}]}$, where ${\cal G}^{[s,{\bf r}]}$ acts on the spin-flavor part of the wave function.
The matrix element in its most general form can be written as
\begin{eqnarray}\label{Sectwo1}
 &&\langle(L,S)J,J_z;{\bf R},Y,I,I_z| \left(\xi^{(l)}  {\cal G}^{[s,{\bf r}]}\right)^{[j,{\bf r}]}|(L',S')J',J'_z;{\bf R}',Y',I',I'_z\rangle \nonumber\\
 \nonumber\\
&=& (-1)^{J'-J'_z}
  \left(
    \begin{array}{cc|c}
        J   	& J' 	& j   \\
        J_z 	& -J'_z & j_{z}
    \end{array}
\right) \sum_\gamma
  \left(
    \begin{array}{cc|c}
        {\bf R}'   	& {\bf r} 	& {\bf R}   \\
        Y',I',I'_z 	& \nu 	& Y,I,I_z
    \end{array}
\right)_\gamma
\frac{1}{\sqrt{D({\bf R})}}\nonumber\\
&\times&  {\hat J}{\hat J'}
 \left\{
    \begin{array}{ccc}
        L   	& L' 		& l   \\
        S 	& S'		& s   \\
        J 	& J'		& j
    \end{array}
\right\}
\langle L|| \xi^{(l)}||L'\rangle
\langle S;{\bf R}|| {\cal G}^{[s,{\bf r}]}||S';{\bf R}'\rangle_\gamma ,
\end{eqnarray}
where ${\hat J}\equiv\sqrt{1+2J}$, $D({\bf R})$ is the dimension of the representation $\bf R$, the second term in parentheses is a Clebsch-Gordan coefficient in $SU(3)$ (defined by Eq.~(\ref{EqAppC1})) and the term in braces is an ordinary $SU(2)$ 9j symbol.
In the cases of mass operators $j=0$ and ${\bf r}={\bf 1}$. The deduction of this expression can be found on App.~\ref{AppC}. Reduced matrix elements of $\xi^{(l)}$ are left undetermined to maintain generality. As described in detail in App.~\ref{AppC}, reduced matrix elements for each ${\cal G}^{[s,{\bf r}]}$ operator can be expressed in terms of reduced matrix elements of core operators whose explicit expressions can be found at the end of that appendix.

States in Eq.~(\ref{states2}) with same spin and isospin can mix giving rise to 24 mass matrices:
8 nonstrange mass matrices $N$, $\Delta$, with $J=1/2,3/2,5/2,7/2$ and 16 matrices $\Sigma$, $\Lambda$, $\Xi$, $\Omega$ with $J=1/2,3/2,5/2,7/2$ containing strange states.
Some constants in the matrix elements can be absorbed in the $c_i^{\bf T,T'}$ coefficients, including reduced matrix elements of $\xi^{(1)}$ and $\xi^{(2)}$ in such way that ${c}_2^{MS_2},{c}_2^{S_2,MS_2},c_4^{MS_2},c_4^{S_2,MS_2}\sim \langle 2 ||\xi^{(1)} || 2 \rangle$, $c_3^{MS_2}\sim \langle 2 ||\xi^{(2)} || 2 \rangle$ and $c_3^{MS_0,MS_2}\sim \langle0 ||\xi^{(2)} || 2 \rangle$.

After a simple inspection of the matrix expressions, we found that a more convenient way of writing the nonstrange matrices is by making the replacements ${\bar c}_1^{\bf T,T'}=c_1^{\bf T,T'}+\frac{1}{N_c}c_5^{\bf T,T'}$ and \mbox{${\bar c}_2^{S_2,MS_2}=c_2^{S_2,MS_2}-c_4^{S_2,MS_2}$, ${\bar c}_2^{MS_2}=c_2^{MS_2}+c_4^{MS_2}$}. Since all mass matrices are symmetric, when presented, we show only the upper right part. To illustrate the nonstrange case we show here the $N_{3/2}$ mass matrix with these redefinitions:
\begin{eqnarray}
 M_{N_{3/2}}=\left(
\begin{array}{cccc}
 {\bar c}_1^{S_2} N_c & 0 & {\bar c}_2^{S_2,MS_2} & -{\bar c}_2^{S_2,MS_2} \\
   & {\bar c}_1^{MS_0} N_c & -c_3^{MS_0,MS_2} & -c_3^{MS_0,MS_2} \\
   &   & {\bar c}_1^{MS_2} N_c-{\bar c}_2^{MS_2} & -\frac{1}{2}{\bar c}_2^{MS_2}-c_3^{MS_2} \\
   &   &   & {\bar c}_1^{MS_2} N_c-{\bar c}_2^{MS_2} \\
\end{array}
\right),
\end{eqnarray}
in the $\left\{[{\bf ``56"},2^+]^{[\frac12,{\bf8}]}, [{\bf ``70"},0^+]^{[\frac32,{\bf8}]}, [{\bf ``70"},2^+]^{[\frac12,{\bf8}]}, [{\bf ``70"},2^+]^{[\frac12,{\bf8}]} \right\}$ basis where the superscripts indicate the spin and flavor representations respectively.
The diagonalization of this matrix leads to four eigenvalues we denote as $m_K$, with $K=1^\pm$,~$2^\pm$, in this case. The $m_K$ are given in terms of  the $c_i^{\bf T,T'}$ coefficients in the next section.
The corresponding eigenvectors are
\begin{eqnarray}\begin{array}{ccl}
N_{3/2}^{K=1^+} &=& (-\eta_{MS_0},0,\sqrt{1/2}, \sqrt{1/2}),\\
N_{3/2}^{K=1^-} &=& (1, 0, \sqrt{1/2}\,\eta_{MS_0}, \sqrt{1/2}\,\eta_{MS_0}),\\
N_{3/2}^{K=2^+} &=& (0, -\eta_{S_2}, -\sqrt{1/2}, \sqrt{1/2}),\\
N_{3/2}^{K=2^-} &=& (0, 1, -\sqrt{1/2}\,\eta_{S_2}, \sqrt{1/2}\,\eta_{S_2}),  \end{array}
\end{eqnarray}
where $\eta_{\,\bf T}$ can be expressed in terms of the $c_i^{\bf T,T'}$ coefficients and will be given explicitly in the next section.
All matrix elements for the nonstrange states can be written in terms of the coefficients ${\bar c}_1^{\bf T,T'}$, ${\bar c}_2^{\bf T,T'}$ and $c_3^{\bf T,T'}$ and were found to have the same expressions as the matrices of Ref.~\cite{Willemyns:2017jnh}.

As an illustrative case for the strange states, we present here the results for the $\Sigma_{7/2}$ states in the \mbox{large $N_c$} limit. The mass matrix in the basis $\left\{[{\bf ``56"},2^+]^{[\frac32,{\bf 10}]}, [{\bf ``70"},2^+]^{[\frac32,{\bf 8}]}, [{\bf ``70"},2^+]^{[\frac32,{\bf S}]}, [{\bf ``70"},2^+]^{[\frac32,{\bf 10}]}, [{\bf ``70"},2^+]^{[\frac52,{\bf 10}]}\right\}$ can be expressed as $M_{\Sigma_{7/2}}=M_{\Sigma_{7/2}}^{LO}+M_{\Sigma_{7/2}}^{NLO}$ where
\begin{eqnarray}
 M_{\Sigma_{7/2}}^{LO}=\mbox{diag}\left( {\bar c}_1^{S_2} , {\bar c}_1^{MS_2} , {\bar c}_1^{MS_2} , {\bar c}_1^{MS_2},{\bar c}_1^{MS_2} \right) N_c\nonumber \ ,
\end{eqnarray}
and the ${\cal O}(N_c^0)$ contribution is given by
\begin{eqnarray}
 M_{\Sigma_{7/2}}^{NLO}=\left({
{\arraycolsep=1.50pt
\begin{array}{ccccc}
 0 & 0 & 0 & -\frac{2}{\sqrt{5}}{\bar c}_2^{S_2,MS_2} & -\sqrt{\frac{6}{5}} {\bar c}_2^{S_2,MS_2} \\
   & {\bar c}_2^{MS_2} -\frac{2}{7} c_3^{MS_2} & 0 & 0 & 0 \\
   &   & {\bar c}_2^{MS_2}-c_4^{MS_2}-3 c_5^{MS_2} & 0 & 0 \\
   &   &   & \frac{2}{5} {\bar c}_2^{MS_2}+\frac{8}{35} c_3^{MS_2} &  \frac{9}{35} \sqrt{6} c_3^{MS_2} - \frac35\sqrt{\frac{3}{2}}{\bar c}_2^{MS_2} \\
   &   &   &   & \frac{1}{10}{\bar c}_2^{MS_2}+\frac{17}{35} c_3^{MS_2}
\end{array}}
}
\right). \nonumber
\end{eqnarray}
In contrast with the nonstrange cases, mass matrices of strange states have contributions from coefficients $c_4^{\bf T,T'}$ and $c_5^{\bf T,T'}$. The $\Sigma_{7/2}$ mass matrix has four eigenvalues $m_K$, with $K=2^+$,~$2^-$,~$3$,~$5/2$.
The corresponding eigenvectors are given by
\begin{eqnarray}
\begin{array}{ccl}
\Sigma_{7/2}^{K=\frac52\,\,\,\,}&=&(0,0,1, 0, 0 ), \nonumber\\
\Sigma_{7/2}^{K=3\,\,\,\,}&=&(0, 0, 0, -\sqrt{3/5}, \sqrt{2/5}), \nonumber\\
\Sigma_{7/2}^{K=3\,\,\,\,}&=&(0, 1, 0, 0, 0), \nonumber\\
\Sigma_{7/2}^{K=2^-}&=&(\eta_{S_2}, 0, 0, -\sqrt{2/5}, -\sqrt{3/5}), \nonumber \\
\Sigma_{7/2}^{K=2^+}&=&(1, 0, 0, \sqrt{2/5}\,\eta_{S_2}, \sqrt{3/5}\,\eta_{S_2}).\nonumber
\end{array}
\end{eqnarray}
Note that there are two states with $K=3$, each one corresponding to states with ${\bf ``8"}$ and ${\bf ``10"}$ flavor symmetry.

As mentioned before, even if the $SU(6)$ symmetry of the quark models does not hold in large $N_c$ QCD, we did not expect all states in the multiplets listed in Eq.~(\ref{states2}) to mix. The spin number $S$ is not a good quantum number in Nature, namely $(I,J)$ states are a linear combination of states with different $S$ number.
Also, since we are neglecting the breaking of the $SU(3)$ symmetry all members of the same flavor multiplets are degenerate and states from different multiplets ${``\bf8"}$, ${``\bf10"}$, ${``\bf1"}$ and ${``\bf S"}$ do not mix.
However, we allowed for configuration mixing to occur so that the states from different $SU(6)$ multiplets mix as well as states with different $L$.

We found that operators ${\cal O}_2=\left(\xi^{(1)} s\right)^{[0,{\bf 1}]}$ and ${\cal O}_4=\frac{1}{N_c}\left(\xi^{(1)} t G_c\right)^{[0,{\bf 1}]} $ allow for the $[{\bf ``56"},2^+]$ and $[{\bf ``70"},2^+]$ multiplets to mix while ${\cal O}_3=\left(\xi^{(2)} g  G_c\right)^{[0,{\bf 1}]}$ allows for  the $[{\bf ``70"},0^+]$ and $[{\bf ``70"},2^+]$ multiplets to mix. Operator ${\cal O}_5=\frac{1}{N_c} \left(t T_c\right)^{[0,{\bf 1}]}$ does not contribute to the mixing of configurations.

By calculating the eigenvalues of the 24 mass matrices we found that all the S and MS states of the ${\cal N}=2$ band have only nine masses which can be expressed as
\begin{eqnarray}\label{masses}
\begin{array}{lcllcl}
 m_0&=&{\bar c}_1^{S_0} N_c,		& m_{\frac12}&=&{\bar c}_1^{MS_0} N_c-3 c_5^{MS_0},\\
 m_{1^\pm}&=&\bar m_{1}\pm\delta_1, 	& m_{\frac32}&=&{\bar c}_1^{MS_2} N_c-\frac{3 }{2}{\bar c}_2^{MS_2}+3 c_4^{MS_2}-3 c_5^{MS_2},\\
 m_{2^\pm}&=&\bar m_{2}\pm\delta_2, 	& m_{\frac52}&=&{\bar c}_1^{MS_2} N_c+{\bar c}_2^{MS_2}-2 c_4^{MS_2}-3 c_5^{MS_2},\\
 m_3&=&{\bar c}_1^{MS_2} N_c+c_2^{MS_2}-\frac{2}{7} c_3^{MS_2}, \quad& & &
\end{array}
\end{eqnarray}
where
\begin{eqnarray}\label{masses2}
 \delta_1&=&\sqrt{\left(\frac12 \left({\bar c}_1^{MS_0}-{\bar c}_1^{MS_2}\right)N_c+\frac34 {\bar c}_2^{MS_2}+\frac12 c_3^{MS_2}\right)^2+2 \left(c_3^{MS_0,MS_2}\right)^2},\nonumber\\
 \delta_2&=&\sqrt{\left(\frac12 \left({\bar c}_1^{S_2}-{\bar c}_1^{MS_2}\right)N_c + \frac14 {\bar c}_2^{MS_2}-\frac12 c_3^{MS_2}\right)^2+2 \left({\bar c}_2^{S_2,MS_2}\right)^2
  },\nonumber
\end{eqnarray}
and
\begin{eqnarray}\label{masses3}
 \bar m_{1}&=&\frac12\left({\bar c}_1^{MS_0}+{\bar c}_1^{MS_2}\right)N_c-\frac34 {\bar c}_2^{MS_2}-\frac12 c_3^{MS_2},\nonumber\\
 \bar m_{2}&=&\frac12 \left({\bar c}_1^{MS_2}+{\bar c}_1^{S_2}\right)N_c -\frac14 {\bar c}_2^{MS_2}+ \frac12 c_3^{MS_2}.\nonumber
\end{eqnarray}

All 24 mass matrices and their eigenvalues can be expressed in terms of only 11 coefficients corresponding to ${\bar c}_1^{S_0}$, ${\bar c}_1^{MS_0}$, ${\bar c}_1^{S_2}$, ${\bar c}_1^{MS_2}$, ${\bar c}_2^{MS_2}$, $c_3^{MS_2}$, $c_4^{MS_2}$, $c_5^{MS_0}$, $c_5^{MS_2}$, ${\bar c}_2^{S_2,MS_2}$, $c_3^{MS_0,MS_2}$. The first nine coefficients are associated to the nine towers while the coefficients ${\bar c}_2^{S_2,MS_2}$ and $c_3^{MS_0,MS_2}$ parametrize the mixing of the spin-flavor multiplets.

When writing the eigenvalues $m_K$ obtained in terms of the mass eigenvalues $\mathring m_K$ that we would have in the absence of configuration mixing, (which is equivalent to setting \mbox{$c_3^{MS_0,MS_2}={\bar c}_2^{S_2,MS_2}=0$}) we find
that $m_K = \mathring m_K$ for $K=0,3$ and for $K=\frac12,\frac32, \frac52$ while for the $K=1,2$ states we obtain the same result as in the $N_f=2$ case, namely
\begin{eqnarray}
m_{K^\pm} &=& \frac{\mathring m_{K^+}+ \mathring m_{K^-}}{2} \pm \sqrt{ \left(\frac{\mathring m_{K^+}-\mathring m_{K^-}}{2}\right)^2 + \left({\mu_K}\right)^2} \ ,
\end{eqnarray}
where $\mu_{1}=-\sqrt{2}\,\,c_3^{MS_0,MS_2}$ and $\mu_{2}=-\sqrt{2}\,\,{\bar c}_2^{S_2,MS_2}$.
With these expressions $ \eta_{MS_0}, \eta_{S_2} $ can be written as
\begin{eqnarray}
\eta_{MS_0}   &=& \frac{2 \mu_1}{\mathring m_{1'}-\mathring m_{1} +
\sqrt{\left(\mathring m_{1'}-\mathring m_{1}\right)^2 + 4 \left(\mu_1\right)^2 } } \ , \nonumber\\
\eta_{S_2}   &=& \frac{2 \mu_2}{\mathring m_{2'}-\mathring m_{2} +
\sqrt{\left(\mathring m_{2'}-\mathring m_{2}\right)^2 + 4 \left(\mu_2\right)^2 } } \ . \nonumber
\end{eqnarray}

The $SU(3)$ multiplets considered organize into nine towers as follows
\begin{eqnarray}\label{Nine1}
  m_0		&:& \left(N_{1/2}, \Lambda^{\bf 8}_{1/2}, \Sigma^{\bf 8}_{1/2}, \Xi^{\bf8}_{1/2}\right), \left(\Delta_{3/2},
		\Sigma^{\bf 10}_{3/2}, \Xi^{\bf10}_{3/2}, \Omega^{\bf10}_{3/2}\right), \nonumber\\
  m_{1^\pm}	&:& \left(N_{1/2}, \Lambda^{\bf 8}_{1/2}, \Sigma^{\bf 8}_{1/2}, \Xi^{\bf8}_{1/2}\right), \left(N_{3/2},
		\Lambda^{\bf 8}_{3/2}, \Sigma^{\bf 8}_{3/2}, \Xi^{\bf8}_{3/2}\right),
		\left(\Delta_{1/2}, \Sigma^{\bf 10}_{1/2}, \Xi^{\bf10}_{1/2}, \Omega^{\bf10}_{1/2}\right),\nonumber\\
		&&  \left(\Delta_{3/2}, \Sigma^{\bf 10}_{3/2}, \Xi^{\bf10}_{3/2}, \Omega^{\bf10}_{3/2}\right), \left(\Delta_{5/2}, \Sigma^{\bf 10}_{5/2}, \Xi^{\bf10}_{5/2}, \Omega^{\bf10}_{5/2}\right), \nonumber\\
  m_{2^\pm}	&:& \left(N_{3/2}, \Lambda^{\bf 8}_{3/2}, \Sigma^{\bf 8}_{3/2}, \Xi^{\bf8}_{3/2}\right),
		\left(N_{5/2}, \Lambda^{\bf 8}_{5/2}, \Sigma^{\bf 8}_{5/2}, \Xi^{\bf8}_{5/2}\right),
		\left(\Delta_{1/2}, \Sigma^{\bf 10}_{1/2}, \Xi^{\bf10}_{1/2}, \Omega^{\bf10}_{1/2}\right),\nonumber\\
		&& \left(\Delta_{3/2}, \Sigma^{\bf 10}_{3/2}, \Xi^{\bf10}_{3/2}, \Omega^{\bf10}_{3/2}\right),
		\left(\Delta_{5/2}, \Sigma^{\bf 10}_{5/2}, \Xi^{\bf10}_{5/2}, \Omega^{\bf10}_{5/2}\right),
		\left(\Delta_{7/2}, \Sigma^{\bf 10}_{7/2}, \Xi^{\bf10}_{7/2}, \Omega^{\bf10}_{7/2}\right), \\
  m_3		&:& \left(N_{5/2}, \Lambda^{\bf 8}_{5/2}, \Sigma^{\bf 8}_{5/2}, \Xi^{\bf8}_{5/2}\right),
		\left(N_{7/2}, \Lambda^{\bf 8}_{7/2},\Sigma^{\bf 8}_{7/2}, \Xi^{\bf8}_{7/2}\right),
		\left(\Delta_{3/2}, \Sigma^{\bf 10}_{3/2}, \Xi^{\bf10}_{3/2}, \Omega^{\bf10}_{3/2}\right),
		\nonumber\\
		&& \left(\Delta_{5/2}, \Sigma^{\bf 10}_{5/2}, \Xi^{\bf10}_{5/2}, \Omega^{\bf10}_{5/2}\right),
		\left(\Delta_{7/2}, \Sigma^{\bf 10}_{7/2}, \Xi^{\bf10}_{7/2}, \Omega^{\bf10}_{7/2}\right), \nonumber\\
  m_\frac12	&:& \left(\Lambda^{\bf 1}_{1/2}, \Xi^{\bf1}_{1/2}\right),
		\left(\Sigma^{\bf S}_{1/2}, \Xi^{\bf S}_{1/2}, \Omega^{\bf S}_{1/2}\right),
		\left(\Sigma^{\bf S}_{3/2}, \Xi^{\bf S}_{3/2}, \Omega^{\bf S}_{3/2}\right), \nonumber\\
  m_\frac32	&:& \left(\Lambda^{\bf 1}_{3/2}, \Xi^{\bf1}_{3/2}\right),
		\left(\Sigma^{\bf S}_{1/2}, \Xi^{\bf S}_{1/2}, \Omega^{\bf S}_{1/2}\right),
		\left(\Sigma^{\bf S}_{3/2}, \Xi^{\bf S}_{3/2}, \Omega^{\bf S}_{3/2}\right),
		\left(\Sigma^{\bf S}_{5/2}, \Xi^{\bf S}_{5/2}, \Omega^{\bf S}_{5/2}\right), \nonumber\\
  m_\frac52	&:& \left(\Lambda^{\bf 1}_{5/2}, \Xi^{\bf1}_{5/2}\right),
		\left(\Sigma^{\bf S}_{3/2}, \Xi^{\bf S}_{3/2}, \Omega^{\bf S}_{3/2}\right),
		\left(\Sigma^{\bf S}_{5/2}, \Xi^{\bf S}_{5/2}, \Omega^{\bf S}_{5/2}\right),
		\left(\Sigma^{\bf S}_{7/2}, \Xi^{\bf S}_{7/2}, \Omega^{\bf S}_{7/2}\right), \nonumber
\end{eqnarray}
where we grouped states by flavor multiplet.
All nonstrange states appear in the $m_0$, $m_{1^\pm}$, $m_{2^\pm}$, $m_3$ towers (and the results are consistent with  Ref.~\cite{Willemyns:2017jnh}) so do their strange multiplet partners.
These expressions for $m_0$, $m_{1^\pm}$, $m_{2^\pm}$, $m_3$ are the same as for the nonstrange case.
On another hand, the $m_{1/2}$, $m_{3/2}$ and $m_{5/2}$ towers contain only $\Lambda\supset``{\bf1}"$ and spurious $\Sigma,\Xi,\Omega\supset``{\bf S}"$ and $\Xi\supset``{\bf 1}"$ states.
The classification of states listed in Eq.~(\ref{Nine1}) reveals a remarkable structure, it indicates that all 146 isomultiplets considered have only nine distinct eigenvalues (in the $SU(3)$ limit).

It is easy to see from Eq.~(\ref{masses}) that nonstrange states can be described by using only operators $O_1$, $O_2$, $O_3$ since in $SU(2)$ subspace $O_5$ is proportional to $O_1$ and $O_4$ is proportional to $O_2$, however the proportionality constant is different when considering the ${\bf T},{\bf T}'=S_2,MS_2$ or the ${\bf T},{\bf T}'=MS_2$ subspace, indicated by the replacements ${\bar c}_2^{S_2,MS_2}=c_2^{S_2,MS_2}-c_4^{S_2,MS_2}$, ${\bar c}_2^{MS_2}=c_2^{MS_2}+c_4^{MS_2}$ we did in Sec.~\ref{Sec4}.

Expressions of matrix elements in the case of finite $N_c$ are long. Thus we limit ourselves only to mention that in all cases spurious states decouple from the physical states in the limit $N_c=3$.

It is interesting at this point to present a brief discussion about the possible determination
of the mass scale of the towers in Eq.(\ref{Nine1}). From purely large $N_c$ considerations, the energy differences between towers inside a band are of order ${\cal O}(N_c^0)$. However, as observed from fits of the $[{\bf70},1^-]$ multiplet (see e.g. Ref.~\cite{Goity:2002pu}), the towers in a given band are closer than expected
from this $1/N_c$ argument (e.g. the spin-orbit operator is particularly small).
This can be seen as a consequence of the fact that, in Nature, resonances respond to an approximated quark model symmetry. Since this same outcome is expected for the band ${\cal N}=2$ multiplets, the assignment of quark model energies
to the large $N_c$ (i.e  ${\cal O}(N_c^0)$) towers is problematic. Namely, to be able to reasonably match the expansion coefficients to some quark model parameters as done in e.g. Ref.~\cite{Semay:2007cv} one needs to go beyond the \mbox{$N_c^0$-approximation} considered in the present work. This requires to consider $1/N_c$ and flavor $SU(3)$ breaking corrections that would break the large $N_c$ towers giving a more similar spectrum as the one obtained from quark models as was done in Ref.~\cite{Willemyns:2015hgy} for the $[{\bf70},1^-]$. It would be particularly interesting to include $1/N_c$ contributions to match the results to quark models that include chiral symmetry breaking following the lines of the mapping performed in Ref.~\cite{Bicudo:2016eeu}.

\section{Towers in $SU(3)$}\label{Sec5}

As mentioned in the Introduction, in the \mbox{large $N_c$} limit a classification of the baryons into towers arises.
As a consequence, when using a generalized quark model basis, only states with the same $K$ value can mix. And compared to the no-mixing case, the configuration mixing only shifts the energies of towers.

For a given state in a $SU(6) \times O(3)$ representation, towers for nonstrange states are given by fairly simple relations between the $K$ number and the orbital angular momentum, i.e., ${\bf K}={\bf L}$ for the symmetric representations and ${\bf K}={\bf L}+{\bf 1}$ for the mixed-symmetric ones \cite{Willemyns:2017jnh}. If there is no symmetry breaking these relations must hold for the strange states in the corresponding $SU(3)$ flavor multiplets.

For states belonging to $SU(3)$ flavor representations $``{\bf S}"$ and $``{\bf 1}"$, which do not have nonstrange states, we associated a half-integer $K$ value which indicates that the $K$ number and the orbital angular momentum relations found for $N_f=2$ do not hold in the $SU(3)$ case in general.

We found that in a $SU(3)$ generalization one can consider the lower strangeness number $n_{s,min}$ of the flavor multiplet. Defining $M=1-\frac{n_{s,min}}{2}$ the $K$ relation for the mixed-symmetric representations is ${\bf K}={\bf L}+{\bf M}$.
In contrast with the $SU(2)$ case, there are no general expressions for $K$ depending only on the $O(3)$ representation, we also need a the flavor representation term.
Since there are no $``{\bf S}"$ or $``{\bf 1}"$ flavor multiplets contained in $\left[``{\bf56}",0\right]$ and $\left[``{\bf56}",2\right]$ the relation ${\bf K}={\bf L}$ still holds for the S representations and their states fall into $K=0$ and $K=2$ towers respectively. It is easy to see with the generalized relation proposed for the MS representations, that states in $\left[``{\bf70}",0\right]$ belonging to $``{\bf 8}"$ or $``{\bf 10}"$ flavor multiplets have $K=1$ while states from the $``{\bf S}"$ or $``{\bf 1}"$ flavor representations have $K=\frac12$. States from $\left[``{\bf70}",2\right]$ with $``{\bf 8}"$ or $``{\bf 10}"$ flavor symmetry have $K=1,2,3$ and states from $``{\bf S}"$ or $``{\bf 1}"$ have $K=\frac32,\frac52$. This is consistent with having two $K=1$ and two $K=2$ values.

\section{Nucleon-meson scattering picture}\label{Sec6}

As discussed in the Introduction, another method to uncover the properties of \mbox{excited} states is to study the scattering processes deduced exclusively from \mbox{large $N_c$}.
The \mbox{compatibility} of the patterns of degeneracy obtained from the \mbox{large $N_c$} quark model and the resonances directly obtained from \mbox{large $N_c$} was shown explicitly in Ref.~\cite{Cohen:2005ct} for the $[{\bf ``70"},1^-]$ multiplet.

In this section we want to show explicitly, in one hand, that the compatibility also holds for the $[{\bf ``56"},0^+]$, $[{\bf ``70"},0^+]$, $[{\bf `` 56"},2^+]$, $[{\bf ``70"},2^+]$ multiplets and, in another hand, that the $K$ values we attributed for the strange states, are consistent with this picture. In particular, we assigned half-integer $K$ values to the $\bf ``S"$ and $\bf ``1"$ representations.

In order to analyze a resonance with $I_s,J_s$ quantum numbers we study the meson-baryon scattering $\phi(S_\phi,{\bf R}_\phi,I_\phi,Y_\phi) + B(S_B,{\bf R}_B,I_B,Y_B)\to\phi'(S_{\phi'},{\bf R}_{\phi'},I_{\phi'},Y_{\phi'}) + B'(S_{B'},{\bf R}_{B'},I_{B'},Y_{B'})$ where $\phi$ and $B$ stand for  meson and baryon respectively. A resonance is a pole in the scattering amplitude at unphysical kinematics.
The phenomenologically relevant cases are the ones with $0^-$ mesons, so we use the spinless meson expression given by~\cite{Cohen:2005ct}
\begin{eqnarray}\label{scat}
\lefteqn{S_{\ell \ell' S_B S_{B'} J_s {\bf R}_s \gamma_s
\gamma'_s I_s Y_s}} \nonumber \\
& = &
(-1)^{\ell-\ell'}
\frac{\sqrt{D({\bf R}_B)D({\bf R}_{B'})}}{D({\bf R}_s)}
\sum_{\stackrel{\scriptstyle I, I' \! , \, Y \in
{\bf 8},}{I'' \in {\bf R}_s}}
(-1)^{I + I' + Y} \hat I''
\left( \begin{array}{cc||c} {\bf R}_B & {\bf 8} & {\bf R}_s \, \gamma_s \\ S_B
\frac{N_c}{3} & I Y & I'' \, Y \! \! + \! \frac{N_c}{3}
\end{array} \right)
\nonumber \\ & & \times
\left( \begin{array}{cc||c} {\bf R}_B & {\bf 8} & {\bf R}_s \, \gamma_s \\ I_B
Y_B & I_\phi Y_\phi & I_s Y_s \end{array} \right)
\left( \begin{array}{cc||c} {\bf R}_{B'} & {\bf 8} & {\bf R}_s \,
\gamma'_s \\ S_{B'}
\frac{N_c}{3} & I' Y & I'' \, Y \! \! + \!
\frac{N_c}{3} \end{array} \right)
\left( \begin{array}{cc||c} {\bf R}_{B'} & {\bf 8} & {\bf R}_s \,
\gamma'_s \\ I_{B'} Y_{B'} & I_{\phi'}
Y_{\phi'} & I_s Y_s \end{array} \right)
\nonumber \\ & & \times
\sum_{K} \hat K
\left\{ \begin{array}{ccc}
K   & I'' & J_s \\
S_B & \ell                & I \end{array} \right\}
\! \left\{ \begin{array}{ccc}
K            & I'' & J_s \\
S_{B'} & \ell'         & I' \end{array} \right\}
\tau^{I I' Y }_{K \ell \ell'} \ ,
\end{eqnarray}
where ${\bf R}_B = (2S_B,\frac{N_c}{2} - S_B)$ is the representation corresponding to baryons in the ground-state $``{\bf 56}"$
with spin $S_B$, for which the nonstrange states have isospin $I_{B} = S_B$ and \mbox{$Y_{B, max}= \frac{N_c}{3}$}.
States with hypercharge $Y_{max}=\frac{N_c}{3}$, $``{\bf 8}"$ and $``{\bf 10}"$ states, decay via $\pi$ and $\eta$ while $``{\bf 1}"$ and $``{\bf S}"$ states which have $Y_{max}=\frac{N_c}{3}-1$ decay via $\bar{K^0}$ and $K^-$. (To see a detailed example on how Eq.~(\ref{scat}) is applied in this context see Ref.~\cite{Cohen:2005yj}.)

A resonant pole appearing in one of the physical amplitudes must appear in at least one reduced amplitude $\tau_{K\ell\ell'}^{II'Y}$. This reduced amplitude contributes in turn to a number of other physical amplitudes.
For a given resonance with $R_s$, $Y_s$ and $I_s$ quantum numbers we find the $\tau_{K\ell\ell'}^{II'Y}$ amplitudes that contain that pole, the only characteristic number associated with the resonance being $K$.

For our purposes, we only need to consider a small set of scattering processes.
Namely, those for which the desired poles can be accessed so that the possible $K$ values associated to a given resonance can be determined.
Following the lines of Ref.~\cite{Cohen:2005ct} all quantum numbers are chosen diagonal
\mbox{($B=B'$, $\phi=\phi'$ , $\ell=\ell'$)} and we analyze only ${\bf ``8"}\to{\bf ``8"}$ transitions except for the cases in which this scattering does not access all poles, in that case we also show the ${\bf ``10"}\to{\bf ``10"}$ transitions.
The obtained results are presented in Tables~\ref{TabAppB1}-\ref{TabAppB7} of App.~\ref{AppB} where we also include the results obtained with the operator method.
We can read that a state found to have a mass $m_i$ in the $1/N_c$ expansion with quark operators can be associated with a resonance that occurs in the $K=i$ scattering channel.
Resonant poles are obtained if the poles are located at the values $m_K$, thus the states described in Eq.~(\ref{states2}) organize as the pattern found on Eq.~(\ref{Nine1})

From tables of App.~\ref{AppB} we observe that some $K$ values found from the resonance picture have not a counterpart in the results using the $1/N_c$ expansion. As noted in Ref.~\cite{Cohen:2003jk}, this does not mean the pictures are not compatible but rather that the higher $K$ amplitudes correspond to resonances of higher orbital momentum (with the same parity) that reveal themselves in the same channel.

It is interesting to note that the consequence of breaking $SU(3)$ symmetry in the resonance picture is that only $\tau$ with different $Y$ become distinct (even if they have the same $K$ number). So the resonance picture suggests that states with same hypercharge in the same $K$ tower will remain degenerate even if $SU(3)$ symmetry is arbitrarily broken.

\section{The antisymmetric case: the $[{\bf ``20"}, 1^+]$ multiplet}\label{Sec7}

As mentioned in Sec.~\ref{Sec2}, in the $N_c>3$ generalization we assume that the additional $N_c-3$ quarks appear in a completely symmetric spin-flavor combination. The denoted antisymmetric representation is only fully antisymmetric for $N_c=3$.
As for the $MS$ and $S$ multiplets, to analyze the masses of antisymmetric states, we need to know the $SU(2)$ and $SU(N_f)$ contents of the $SU(2N_f)$ states.
Using the general method described in Ref.~\cite{Hagen:1965} we obtained the $SU(6)$ decomposition into spin and flavor representations in the antisymmetric case:
\begin{eqnarray}\label{states1}
{\rm A} \equiv (N_c \! - \! 3, 0, 1, 0, \ldots, 0) & = &
\bigoplus_{n=0}^{(N_c-5)/2} \left[n + \frac32 , \left( 2n+1, \frac 1 2 (N_c-1) - n, 0, 0, 0, \ldots , 0 \right) \right] \nonumber \\ & &
\bigoplus_{n=0}^{(N_c-7)/2} \left[n + \frac12 , \left( 2n+1, \frac 1 2 (N_c-7) - n, 2, 0, 0, \ldots , 0 \right) \right] \nonumber \\ & &
\bigoplus_{n=0}^{(N_c-5)/2} {\mbox{\Large2 }} \left[n + \frac12, \left( 2n+2, \frac 1 2 (N_c-5) - n, 1, 0, \ldots , 0 \right) \right] \nonumber \\ & &
\bigoplus_{n=0}^{(N_c-5)/2} \left[n + \frac12,  \left( 2n+3, \frac 1 2 (N_c-3) - n, 0, 0, \ldots , 0 \right) \right] \nonumber \\ & &
\bigoplus_{n=0}^{(N_c-3)/2} \left[n + \frac12, \left( 2n+1, \frac 1 2 (N_c-1) - n, 0, 0, \ldots , 0 \right) \right]  \nonumber \\ & &
\bigoplus_{n=0}^{(N_c-7)/2} \left[n + \frac12, \left( 2n+4, \frac 1 2 (N_c-7) - n, 1, 0, \ldots , 0 \right) \right]  \nonumber \\ & &
\bigoplus_{n=0}^{(N_c-5)/2} \left[n + \frac32, \left( 2n+2, \frac 1 2 (N_c-5) - n, 1, 0, \ldots , 0 \right) \right]  \nonumber \\ & &
\bigoplus_{n=0}^{(N_c-5)/2} \left[n + \frac12, \left( 2n, \frac 1 2 (N_c-3) - n, 1, 0, \ldots , 0 \right) \right]  \nonumber \\ & &
\bigoplus_{n=0}^{(N_c-3)/2} \left[n + \frac32, \left( 2n, \frac 1 2 (N_c-3) - n, 1, 0, \ldots , 0 \right) \right].
\end{eqnarray} 
For low values of $N_c$, some terms in Eq.~(\ref{states1}) vanish as the sum labels have to be non-negative and the $n_r$ entries in the Dynkin symbols have to be null when $r>N_f-1$ (or else we would need more flavors to obtain the wanted symmetry), any term that does not fulfill these conditions vanishes.
It is worth noting that according to Eq.~(\ref{states1}) the first three terms in equation (3.3) of Ref.~\cite{Cohen:2003nn} should not be present.

After a straightforward examination of the states of each multiplet, we find that the SU(6) decompositions useful to this analysis for each spin-flavor multiplet are given by
\begin{eqnarray}
  \left[``{\bf20}",1\right]:& & \left[1/2,``{\bf8}"\right]^{1/2} \oplus \left[3/2,``{\bf8}"\right]^{1/2} \oplus \left[1/2,``{\bf1}"\right]^{3/2} \oplus \left[3/2,``{\bf1}"\right]^{3/2} \oplus \nonumber\\
 && \left[5/2,``{\bf1}"\right]^{3/2} \oplus \left[1/2,``{\bf8}"\right]^{3/2*} \oplus \left[3/2,``{\bf8}"\right]^{3/2*} \oplus \left[5/2,``{\bf8}"\right]^{3/2*} \oplus \\
 && \left[1/2,``{\bf1}"\right]^{1/2*} \oplus \left[3/2,``{\bf1}"\right]^{1/2*} \oplus \dots\nonumber  .
\end{eqnarray}
There are two physical octets with $J=1/2,3/2$ and three singlet $\Lambda$ states with \mbox{$J=1/2,3/2,5/2$} associated with baryons expected to appear in Nature.

\subsection{Antisymmetric wave functions}
To build the antisymmetric wave functions we assume that the cores are mixed-symmetric states of $N_c-1$ quarks. Antisymmetric states are, therefore, a linear combination of mixed-symmetric cores coupled to a quark which can be written as
\begin{equation}\label{statesanti2}
 \left.\big| S, {\bf R}\right\rangle_{A} = \sum_{i} c_i \left.\big| ([S_{c_i},{\bf R}_{c_i}]_{MS}\,\,\,{\bf q})^{[S,{\bf R}]} \right\rangle,
\end{equation}
where  ${\bf q}{\equiv}[1/2,{\bf 3}]$ represents the single quark and the MS label in the core representation indicates that the core has mixed symmetry in the spin-flavor space. The MS core and the excited quark  quantum numbers are coupled in such way that the overall symmetry corresponds to the spin and flavor representations $S$ and ${\bf R}$.  The coefficients $c_i$ in Eq.~(\ref{statesanti2}) have to be set to those that give the antisymmetric spin-flavor representation.

The MS cores, in turn, are a combination of a symmetric state of $N_c-2$ quarks coupled to a quark which can be written as
\begin{equation}\label{coreswf}
 \left.\big|S_c, {\bf R}_c\right\rangle_{MS} = \sum_{j} d_j \left.\big| ([S_{\tilde c_j},{\bf R}_{\tilde c_j}]_{S}\,\,\,{\bf q})^{[S_c,{\bf R}_c]} \right\rangle,
\end{equation}
where $d_j$ are known coefficients (given by the $c_{MS}$ of Eq.~(\ref{states5}) with the replacement \mbox{$N_c \rightarrow N_c-1$}).

The found decomposition of the mixed-symmetric cores for arbitrary $N_c$ and $N_f$ is given by
\begin{eqnarray}
{\rm MS_{core}} \equiv (N_c \! - \! 3, 1, 0, 0, \ldots, 0) & = &
\bigoplus_{n=0}^{(N_c-3)/2} \left[n , \left( 2n+2, \frac 1 2 (N_c-3) - n, 0, 0, 0, \ldots , 0 \right) \right] \nonumber \\ & &
\bigoplus_{n=0}^{(N_c-5)/2} \left[n + 1 , \left( 2n+2, \frac 1 2 (N_c-3) - n, 0, 0, 0, \ldots , 0 \right) \right] \nonumber \\ & &
\bigoplus_{n=0}^{(N_c-3)/2} \left[n + 1 , \left( 2n, \frac 1 2 (N_c-1) - n, 0, 0, \ldots , 0 \right) \right]  \\ & &
\bigoplus_{n=0}^{(N_c-5)/2} \left[n , \left( 2n+1, \frac 1 2 (N_c-5) - n, 1, 0, \ldots , 0 \right) \right] \nonumber \\ & &
\bigoplus_{n=0}^{(N_c-5)/2} \left[n + 1 , \left( 2n+1, \frac 1 2 (N_c-5) - n, 1, 0, \ldots , 0 \right) \right] \nonumber. \label{eqAcores}
\end{eqnarray}

At this point, the only parameter left to determine in order to express the antisymmetric waves functions in the uncoupled basis are the $c_i$ coefficients of Eq.~(\ref{statesanti2}).

To find the appropriate linear combination of MS cores we use the quadratic Casimir operator.
The $SU(6)$ quadratic Casimir operator, whose matrix elements are known, can be broken down into $SU(2)$ and $SU(3)$ core and excited quark contributions. By calculating the matrix elements of these contributions, a matrix of the Casimir operator can be obtained which, when diagonalized, will give the core composition of the states with definite symmetry. Details of these calculations can be found on App.~\ref{AppE}.
The core composition for the antisymmetric representations of interest in this work is given by
\begin{eqnarray}\label{statesanti3}
\begin{array}{lcl}
 \left.\bigg|\frac{1}{2},{\bf ``8"} \right\rangle_{A} &=&
 -\frac{1}{\sqrt{2}} \left.\bigg|\left(\left[1,{\bf R}_0\right]_{MS}\,{\bf q}\right)^{\left[\frac{1}{2},{\bf 8}\right]}\right\rangle
 +\frac{1}{\sqrt{2}} \left.\bigg|\left(\left[0,{\bf R}_2\right]_{MS}\,{\bf q}\right)^{\left[\frac{1}{2},{\bf 8}\right]}\right\rangle, \\ \\
 \left.\bigg|\frac{3}{2},{\bf ``8"}\right\rangle_{A} &=&
 -\frac{1}{2\sqrt{2}}\sqrt{\frac{N_c-3}{N_c}} \left.\bigg|\left(\left[1,{\bf R}_0\right]_{MS}\,{\bf q}\right)^{\left[\frac{3}{2},{\bf 8}\right]}\right\rangle  -\frac{3}{4}\sqrt{\frac{N_c-1}{N_c}} \left.\bigg|\left(\left[1,{\bf R}_2\right]_{MS}\,{\bf q}\right)^{\left[\frac{3}{2},{\bf 8}\right]}\right\rangle  \\
 && +\frac{\sqrt{5}}{4}{\sqrt{\frac{N_c+3}{N_c}}} \left.\bigg|\left(\left[2,{\bf R}_2\right]_{MS}\,{\bf q}\right)^{\left[\frac{3}{2},{\bf 8}\right]}\right\rangle , \\ \\
 \left.\bigg|\frac{3}{2},{\bf ``1"} \right\rangle_{A} &=&
 -\frac{1}{\sqrt{2}}\sqrt{\frac{N_c+1}{N_c-1}} \left.\bigg|\left(\left[1,{\bf R}_0\right]_{MS}\,{\bf q}\right)^{\left[\frac{3}{2},{\bf 1}\right]}\right\rangle+\frac{1}{\sqrt{2}}\sqrt{\frac{N_c-3}{N_c-1}} \left.\bigg|\left(\left[1,{\bf R}_1\right]_{MS}\,{\bf q}\right)^{\left[\frac{3}{2},{\bf 1}\right]}\right\rangle ,\\ \\
 \left.\bigg|\frac{1}{2} ,{\bf ``1"}\right\rangle_{A} &=&
 \frac{1}{\sqrt{2}}\sqrt{\frac{(N_c-3)(N_c+1)}{N_c(N_c-1)}} \left.\bigg|\left(\left[1,{\bf R}_0\right]_{MS}\,{\bf q}\right)^{\left[\frac{1}{2},{\bf 1}\right]}\right\rangle -\frac{1}{2}\sqrt{\frac{3}{2}}\sqrt{\frac{N_c-1}{N_c}} \left.\bigg|\left(\left[0,{\bf R}_1\right]_{MS}\,{\bf q}\right)^{\left[\frac{1}{2},{\bf 1}\right]}\right\rangle   \\
 && +\frac{1}{2\sqrt{2}}\frac{N_c+3}{\sqrt{N_c(N_c-1)}} \left.\bigg|\left(\left[1,{\bf R}_1\right]_{MS}\,{\bf q}\right)^{\left[\frac{1}{2},{\bf 1}\right]}\right\rangle ,
\end{array}
\end{eqnarray}
where ${\bf R}_h\equiv\left(h,\frac{N_c-1-7h+3h^2}{2}\right)$.
These coefficients have been checked against those on Ref.~\cite{So:1979nw} for $N_c=5$.

\subsection{Operator expansion}

Since we are not considering $SU(3)$ symmetry breaking, the energy spectrum for \mbox{large $N_c$} for all states belonging to $``{\bf 8}"$ flavor multiplets will be given by the $K$ value that follows from
the ${\bf K}={\bf I}+{\bf J}$ relation for nonstrange states presented in the Introduction.
This implies that there are three towers with $K=0,1,2$.
Then, only three operators are needed to describe the states of flavor multiplets that contain nonstrange baryons of the $[{\bf ``20"},1^+]$ multiplet. The mass operators for the expansion at order ${\cal O}(N_c^0)$ can be chosen to be
\begin{eqnarray}\label{opsanti}
 O_1 &=& \left(N_c \openone\right)^{[0,{\bf1}]},	\nonumber\\
 O_2&=&\left(\xi^{(1)} s\right)^{[0,{\bf 1}]}, \\
 O_3&=&\frac{1}{N_c}\left(\xi^{(2)}\left(g G_c\right)^{[2,{\bf 1}]}\right)^{[0,{\bf 1}]}\nonumber.
\end{eqnarray}
The mass operator can be written as a linear combination of these operators as in Eq.~(\ref{operators1}) where the sum goes up to $i=3$.

The building blocks of the mass operators are, as in the S and MS cases, the $SU(6)$ generators acting on the excited quark and on the core. This clearly implies that the matrix elements of these operators between states containing cores of different symmetry vanish.
Therefore, as mentioned in the Introduction, there is no mixing between baryons of the $[{\bf ``20"},1^+]$ and the other states of the ${\cal N}=2$ band.


Using the operators of Eq.~(\ref{opsanti}) for the nucleon with $J=1/2$, we obtain
\begin{eqnarray}
{\bf M}_{N_{1/2}}=\left(
\begin{array}{cc}
 c_1 N_c-\frac23c_2 & -\frac{1}{3\sqrt{2}}c_2-\frac{5}{24\sqrt{2}}c_3\\
  & c_1 N_c-\frac{5}{6} c_2-\frac{5}{48}c_3 \\
\end{array}
\right).
\end{eqnarray}
This matrix has two eigenvalues that we label $m_0$, $m_1$.

For the $N_{3/2}$, we have
\begin{eqnarray}
{\bf M}_{N_{3/2}}&=&\left(
\begin{array}{cc}
 c_1 N_c+\frac13c_2 & -\frac{\sqrt{5}}{6}c_2-\frac{\sqrt{5}}{48}c_3\\
  & c_1 N_c-\frac{5}{6} c_2-\frac{5}{48}c_3 \\
\end{array}
\right),
\end{eqnarray}
with eigenvalues $m_1$, $m_2$ and
\begin{eqnarray}
{\bf M}_{N_{5/2}}&=&c_1 N_c+\frac12c_2-\frac{1}{48}c_3 \ .
\end{eqnarray}
for the $N_{5/2}$ state which we call $m_3$.

Eigenvalues found for the strange partners in these flavor multiplets are the same as the ones found for the corresponding nucleon.

The eigenvalues expressions found in terms of the expansion coefficients are
\begin{eqnarray}
 m_0&=&c_1 N_c-c_2-\frac{5}{24} c_3 \ ,\nonumber\\
 m_1&=&c_1 N_c-\frac12 c_2+\frac{5}{48}c_3 \ ,\nonumber\\
 m_2&=&c_1 N_c+\frac12 c_2-\frac{1}{48}c_3 \ .\nonumber
\end{eqnarray}


The tower structure found for the nonstrange antisymmetric states and their strange partners in the flavor multiplet is given by
\begin{eqnarray}
  m_0		&:& \left(N_{1/2}, \Lambda^{\bf 8}_{1/2}, \Sigma^{\bf 8}_{1/2}, \Xi^{\bf 8}_{1/2}\right),\nonumber\\
  m_1		&:& \left(N_{1/2}, \Lambda^{\bf 8}_{1/2}, \Sigma^{\bf 8}_{1/2}, \Xi^{\bf8}_{1/2}\right), \left(N_{3/2}, \Lambda^{\bf 8}_{3/2}, \Sigma^{\bf 8}_{3/2}, \Xi^{\bf8}_{3/2}\right),
\label{as}  \\
  m_2		&:& \left(N_{3/2}, \Lambda^{\bf 8}_{3/2}, \Sigma^{\bf 8}_{3/2}, \Xi^{\bf8}_{3/2}\right), \left(N^*_{5/2}, \Lambda^{*(\bf 8)}_{5/2}, \Sigma^{*(\bf 8)}_{5/2}, \Xi^{*(\bf8)}_{5/2}\right)
\nonumber.
\end{eqnarray}

This structure  is the same as the one found for the $[{\bf ``70"},1^-]$ multiplet at \mbox{large $N_c$} where the spin-flavor multiplets containing nonstrange states organize into three towers with $K=0,1,2$ \cite{Cohen:2005ct} and it also agrees with the remark in Ref.~\cite{Cohen:2003nn}, where using a hedgehog-based analysis  authors argue that the MS and A configurations have the
same spectrum of nonstrange states for \mbox{large $N_c$}. Furthermore, not only the tower structure coincides but also the $m_i$ and the matrices expressions are identical to the ones found for the MS states which can be obtained from expressions in Refs.~\cite{Pirjol:2003ye,Carlson:1998vx} (in particular matrices in this work are identical to those in Ref.~\cite{Cohen:2003jk}). Then, the matrices found  imply, as in the MS case, that the mixing angle in the unitary matrix that diagonalizes ${\bf M}_{N_{1/2}}$ and ${\bf M}_{N_{3/2}}$ is surprisingly simple as it is independent of the $c_i$ coefficients. Given the fact that we only have three operators involved and two matrices it is not clear if this is a coincidence or if it suggests that there is a deeper connection between the MS and A symmetries.

\subsection{Towers for antisymmetric states}

We corroborated that multiplets of $[{\bf ``20"},1^+]$ containing nonstrange baryons organize as the MS states of  the $[{\bf ``70"},1^-]$;  the states fall into three towers labeled with $K=0,1,2$. The \mbox{large $N_c$} spectrum of the strange flavor multiplets appear to have a different structure. The states content does no longer match the MS case ($[{\bf ``70"},1^-]$ has only $\left[1/2,``{\bf1}"\right]^{1/2}$, $\left[3/2,``{\bf1}"\right]^{1/2}$). In App.~\ref{AppB2} we present the partial amplitudes containing resonances with quantum numbers corresponding to A states. As can be deduced from Tab.~\ref{TabAppB10}, the resonance picture indicates that, baryons contained in the $[{\bf ``20"},1^+]$ representation belonging to $\left[1/2,``{\bf1}"\right]^{1/2}$, $\left[1/2,``{\bf1}"\right]^{3/2}$, $\left[3/2,``{\bf1}"\right]^{1/2}$, $\left[3/2,``{\bf1}"\right]^{3/2}$, $\left[5/2,``{\bf1}"\right]^{3/2}$ multiplets organize as follows
\begin{eqnarray}
  K={\frac12}	&:& \left(\Lambda^{\bf 1}_{1/2}, \Xi^{\bf1}_{1/2}\right), \nonumber\\
  K={\frac32}	&:& \left(\Lambda^{\bf 1}_{3/2}, \Xi^{\bf1}_{3/2}\right),\\
  K={\frac52}	&:& \left(\Lambda^{\bf 1}_{5/2}, \Xi^{\bf1}_{5/2}\right). \nonumber
\end{eqnarray}
But, for \mbox{large $N_c$} there are two $\Lambda^{\bf 1}_{1/2}$ states with intrinsic spin $1/2$ and $3/2$ that mix resulting in two $\Lambda^{\bf 1}_{1/2}$ eigenstates of the \mbox{\mbox{large $N_c$} QCD}. When considering the resonance picture, only one state with a given $J$ can be assigned to a given energy, the other state has to be identified to a different energy level of the same $K$. Then, in the case of $\Lambda^{\bf 1}_{1/2}$ there are two towers $K={\frac12}$ and $K=\Tilde{\frac12}$. There is an analogous situation for the $\Lambda^{\bf 1}_{3/2}$. When extending the analysis to all nonstrange states of $[{\bf ``20"},1^+]$ five towers seem to appear in addition to the ones containing nonstrange states, namely $K=\frac12,\Tilde{\frac12},\frac32,\Tilde{\frac32},\frac52$. Nevertheless, at this point we turn the attention to the phenomenological relevance such analysis, two towers seem to arise with a same given $K$ value (these towers with same $K$ value should not be confused with the ones of the case of the S and MS states described in previous sections where $K$ towers with $\pm$ labels arise exclusively from configuration mixing) but only one of these two towers contains the physical state, the other is a  ``spurious tower", i.e., this entire tower decouples in the physical limit. Then, it is clear that, from a phenomenological point of view there is no interest in including these two extra spurious towers. Only three non-spurious towers arise which indicates that to  consider these states in a $1/N_c$ expansion framework one should considerate three extra operators in addition to those on Eq.~(\ref{opsanti}).

When building the mass operator for antisymmetric states, reduction rules of Ref.~\cite{Dashen:1994qi} cannot be used for core operators since cores are no longer symmetric. However, the rules apply to the inner core operators since they are $(N_c-2)$-quarks symmetric cores.
Core operators can be decomposed as a sum of an $(N_c-2)$-quarks core operator plus a single quark operator $\Lambda_c=\Lambda_{\tilde c}+\tilde\lambda$, the reduction rules can then be applied to the $\Lambda_{\tilde c}$ operators. On another hand, using the Casimir invariant for the antisymmetric representation given by Eq.~(\ref{casimir2}), and the Casimir invariant for the mixed-symmetric representation with $(N_c-1)$-quarks and for the fundamental representation of a single quark given by
\mbox{$C_{SU(6)}(N_c-3,1,0,0,0)=\frac{1}{12} (N_c-1) (5 N_c+13)$} and \mbox{$C_{SU(6)}(1,0,0,0,0)=\frac{35}{12}$}
respectively, we can express the quadratic Casimir identity for the antisymmetric representation as
\begin{equation}
\frac{2}{3}sS_c+tT_c+4gG_c=-\frac{N_c+11}{6}.
\end{equation}
Then, as in the case of the $\bf ``56"$ and $\bf ``70"$ multiplets, the $gG_c$ operator can always be eliminated in favor of $sS_c$ and $tT_c$.

Considering the reduction rules for symmetric representations applied as described above and the Casimir identity found for A states with MS cores, additional operators  \mbox{$O_4=\frac{1}{N_c}\left(\xi^{(1)}\left(t G_c\right)^{[1,{\bf 1}]}\right)^{[0,{\bf 1}]}$} and
$O_5=\frac{1}{N_c}\left(t T_c\right)^{[0,{\bf 1}]}$ seem to be a good choice of basis for an operator expansion that includes all nonstrange states as they have been used in $[{\bf ``70"},1^-]$ analysis in the $N_f=3$ case. The extra operator to consider could be
$O_6=\frac{1}{N_c}\left(\xi^{(1)}\left(g T_c\right)^{[1,{\bf 1}]}\right)^{[0,{\bf 1}]}$ which in the case of antisymmetric states is not linearly dependent of the other operators since the cores are mixed-symmetric. With these operators one can take a phenomenological approach and calculate the spectra for finite $N_c$ once empirical data about these states become available.

Even if there is no data about the strange antisymmetric states, as mentioned in the Introduction, the nonstrange resonances $N(2100)1/2+$ and $N(2040)3/2+$ have been
tentatively assigned to the antisymmetric multiplet \cite{Crede:2013sze}. In fact, there is a continuing experimental effort in establishing the actual existence of these states.
In particular, in Ref.~\cite{Wang:2017tpe} authors find that they are necessary to describe the cross-section data. Moreover, very recently $N(2100)1/2+$ has been upgraded
from one to three stars in the 2018 edition of Ref.~\cite{Tanabashi:2018}.
In the large $N_c$ picture, the $N(2100)1/2+$ would be associated as the physical state arising from the mixture of $N_{1/2}$ states belonging to the $K=0,1$ towers in Eq.~(\ref{as})
(note that the other combination turns out be unphysical for $N_c=3$).  Similarly, $N(2040)3/2+$ would be associated to the physical mixture of  $N_{3/2}$ states belonging to the $K=1,2$ towers.

\section{Conclusions}\label{Sec8}

In this work we have performed a complete large $N_c$ analysis of the masses of all states belonging to the ${\cal N}=2$ quark model band for $N_c=3$.

We first studied baryons of the ${\cal N}=2$ band in multiplets $[{\bf ``56"},L^+]$ and $[{\bf ``70"},L^+]$ in the \mbox{large $N_c$} limit allowing for the states belonging to the irreducible representation \mbox{$SU(6)\times O(3)$} to mix. This representation arises from the quark models but is not a symmetry of QCD, nor is it a symmetry of \mbox{\mbox{large $N_c$} QCD}.
To analyze the spectrum of these baryons in the \mbox{large $N_c$} limit we considered a $1/N_c$ expansion using core and excited quark operators with a generic spatial operator which allows for the quark model states to mix.
We found that configuration mixing effects appear only on flavor multiplets containing nonstrange states, it has no effect over states in the $``{\bf1}"$ or $``{\bf S}"$ flavor representations.
The 146 isomultiplets of the S and MS representations fall into only nine towers predicted by \mbox{\mbox{large $N_c$} QCD}.
We found that only the $SU(6)\times O(3)$ states with the same $K$ label can mix, which is a direct consequence of the contracted symmetry of the  \mbox{large $N_c$} limit.
Multiplets with nonstrange baryons were found to belong to five towers labeled with $K=0,1^\pm,2^\pm,3$ while strange flavor multiplets were associated with $K=\frac12,\frac32,\frac52$ towers.
We generalized the relation between the $\bf K$ and $\bf L$ numbers for MS states in the $N_f=3$ case finding that it can not be stated as simply as in the nonstrange case since it acquires a dependence on the flavor representation.
In addition to the operator analysis, we showed explicitly that the compatibility of this method and the resonance picture holds for the entire $[{\bf ``56"},L^+]$ and $[{\bf ``70"},L^+]$ multiplets and in particular we showed it still holds when considering configuration mixing.

Using a similar $1/N_c$ expansion with effective quark operators (with a smaller basis), we also showed that $[{\bf ``20"},1^+]$ configurations from flavor multiplets containing nonstrange states also fall into the towers predicted by the \mbox{large $N_c$} symmetries. In addition, we found that the resonance picture gives a compatible classification.
Our results explicitly show that the A states have the
same spectrum of nonstrange states than the MS with \mbox{$L=1$} in the large $N_c$ limit. Furthermore, we observed that nonstrange mass matrices belonging to $[{\bf ``20"},1^+]$ are identical to those from $[{\bf ``70"},1^-]$.
The tower structure was expected to be the same but, considering the nontrivial building of the antisymmetric wave functions, the result that every matrix element is proportional to the MS case was not obvious.
Given the content of these multiplets we could only compare two matrices, it would be very interesting to test this for higher angular momentum states, where there will be more matrices to compare, to further understand if this effect in an overview of a more profound relation between the A and MS representations.
We assumed A states can be described as a MS core coupled to a excited quarks and we found the nontrivial spin-flavor composition of the states belonging to the $[{\bf ``20"},1^+]$ multiplet. This building of the A states gives results compatible with the \mbox{large $N_c$} predictions suggesting that effects from more complex constructions are $N_c$ suppressed.
The baryon-meson scattering picture indicates that the strange multiplets of  $[{\bf ``20"},1^+]$ fall into towers with $K=\frac12,\frac32,\frac52$. However, as explained in Sec.~\ref{Sec7}, there are three towers containing one physical baryon each and two additional spurious towers we labeled with $K=\Tilde{\frac12}, K=\Tilde{\frac32}$.

Using core and excited quark operators we obtained results predicted in \mbox{large $N_c$ QCD} and that are compatible with the resonance picture, even when including configuration mixing.
This indicates that this approach is appropriate to analyze states in the \mbox{large $N_c$} limit and effects from other operators must be subleading.

\section*{Acknowledgments}

We are grateful to C. Schat for useful comments. This work has been supported
in part by CONICET and ANPCyT (Argentina), under grants PIP14-578 and
PICT14-03-0492.


\appendix
\section{Conventions and details of the calculation of relevant matrix elements}\label{AppC}

In addition to the usual definition of reduced matrix element for $SU(2)$ (in this paper we follow conventions of Ref.~\cite{Edmonds:1955fi}) we also use the Wigner-Eckart theorem for $SU(3)$ which is given by
\begin{eqnarray}\label{EqAppC1}
&&\langle {\bf R},\{Y,I,I_z\}| T^{\bf r}_{\{Y^{op},I^{op},I^{op}_z\}}|{\bf R}',\{Y',I',I'_z\}\rangle \nonumber\\
&=&\sum_\gamma
\left(
    \begin{array}{cc|c}
        {\bf R}'   	& {\bf r}	& {\bf R}   \\
        \{Y',I',I'_z\} 	& \{Y^{op},I^{op},I^{op}_z\}	& \{Y,I,I_z\}
    \end{array}
\right)_\gamma
\frac{1}{\sqrt{D({\bf R})}}\langle {\bf R}' ||T^{{\bf r}}||{\bf R}'\rangle_\gamma \ ,
\end{eqnarray}
where $D({\bf R})$ is the dimension of the representation and the parentheses indicate a \mbox{Clebsch-Gordan} coefficient for $SU(3)$.

The Wigner-Eckart theorem in both its $SU(2)$ and $SU(3)$ versions leads to the general expression for the mass matrix elements
\begin{eqnarray}\label{AppC1}
 \langle(L,S)J,J_z;{\bf R},Y,I,I_z| \left(\xi^{(l)} {\cal G}^{[s,{\bf r}]}\right)^{[j,{\bf r}]}|(L',S')J',J'_z;{\bf R}',Y',I',I'_z\rangle \nonumber\\
 \nonumber\\
= (-1)^{J'-J'_z}
  \left(
    \begin{array}{cc|c}
        J   	& J' 	& j   \\
        J_z 	& -J'_z & j_{z}
    \end{array}
\right)
\frac{1}{\hat j} \sum_\gamma
  \left(
    \begin{array}{cc|c}
        {\bf R}'   	& {\bf r} 	& {\bf R}   \\
        Y',I',I'_z 	& \nu 	& Y,I,I_z
    \end{array}
\right)_\gamma
\frac{1}{\sqrt{D({\bf R})}}\\
\times \langle(L,S)J;{\bf R}|| \left(\xi^{(l)}   {\cal G}^{[s,{\bf r}]}\right)^{[j,{\bf r}]}||(L',S')J';{\bf R}'\rangle_\gamma \nonumber \ ,
\end{eqnarray}
where ${\hat j}\equiv\sqrt{1+2j}$.
In the cases of mass operators we will have $j=0$ and ${\bf r}={\bf 1}$.
The orbital and spin-flavor contributions in the reduced matrix element above can be written in an uncoupled basis as
\begin{eqnarray}\label{AppC2}
&&\langle(L,S)J;{\bf R}|| \left(\xi^{(l)}   {\cal G}^{[s,{\bf r}]}\right)^{[j,{\bf r}]}||(L',S')J';{\bf R}'\rangle_\gamma\nonumber\\
&=&  {\hat j}{\hat J}{\hat J'}
 \left\{
    \begin{array}{ccc}
        L   	& L' 	& l   \\
        S 	& S'	& s   \\
        J 	& J'	& j
    \end{array}
\right\}
\langle L|| \xi^{(l)}||L'\rangle
\langle S,{\bf R}|| {\cal G}^{[s,{\bf r}]}||S',{\bf R}'\rangle_\gamma \ ,
\end{eqnarray}
where the term in braces is an ordinary $SU(2)$ 9j symbol. A simple replacement of expression in Eq.~(\ref{AppC2}) into Eq.~(\ref{AppC1}) gives Eq.~(\ref{Sectwo1}).
The terms $\langle L|| \xi^{(l)}||L'\rangle$ are left undetermined in this paper to maintain the generality over the orbital operator.
Then, the terms to determine are the reduced matrix elements $\langle S,{\bf R}|| {\cal G}^{[s,{\bf r}]}||S,{\bf R}'\rangle_\gamma$.

The ${\cal G}$ operator can be written as
\begin{equation}
 {\cal G}^{[s,{\bf r}]} = \left(\lambda^{[s_a,{\bf r_a}]} {\Lambda}^{[s_b,{\bf r_b}]}\right)^{[s,{\bf r}],\gamma_2},
\end{equation}
where $\lambda$ and $\Lambda$ are quark and core operators respectively, $\lambda = s, t, g$ and $\Lambda = S_c,T_c,G_c$. Then, the matrix elements of $\cal G$ can be written in the uncoupled basis as
\begin{eqnarray}\label{EqAppC5}
\langle\left([S_c,{\bf R}_c]\,{\bf q}\right)^{[S, {\bf R}]}|| \left(\lambda^{[s_a,{\bf r}_a]} {\Lambda}^{[s_b,{\bf r}_b]}\right)^{[s,{\bf r}],\gamma_2}||\left([S'_c, {\bf R}'_c]\,{\bf q}\right)^{[S', {\bf R}']}\rangle_\gamma \nonumber\\
= (-1)^{s_a+s_b-s}  \frac{\hat s \hat S \hat S'}{\sqrt{3 D({\bf R})D({\bf R_c})}}
 \left\{
    \begin{array}{ccc}
        S_c	& S'_c	& s_b   \\
        1/2	& 1/2	& s_a   \\
        S	& S'	& s
    \end{array}
\right\}
\sum_{\gamma_a}
\left\{
    \begin{array}{ccc}
        {\bf R}'_c	& {\bf r}_b		& {\bf R}_c,\gamma_a   \\
        {\bf 3}		& {\bf r}_a		& {\bf 3}   \\
        {\bf R}'	& {\bf r},\gamma_2	& {\bf R}
    \end{array}
\right\}_\gamma \nonumber\\
\times \langle S_c,{\bf R}_c ||\Lambda^{[s_b,{\bf r}_b]}|| S'_c,{\bf R}'_c \rangle_{\gamma_a}
\langle{\bf q}||\lambda^{[s_b,{\bf r}_b]}||{\bf q}\rangle,
\end{eqnarray}
where the second term in braces represents a $SU(3)$ 9j symbol which is defined in the following.

The $SU(3)$ 9j symbols are defined by the reduced matrix element of a two-body operator in the $SU(3)$ space written as
\begin{eqnarray}
 &&\langle \left({\bf R}_1,{\bf R}_2\right)^{{\bf R},\gamma}||\left(T_1^{{\bf r}_a}T_2^{{\bf r}_b}\right)^{{\bf r},\alpha}||\left({\bf R}'_1,{\bf R}'_2\right)^{{\bf R}',\gamma'}\rangle_\beta  \\
 &&=
 \frac{1}{\sqrt{D({\bf R}) D({\bf R}_1) D({\bf R}_2)}} \sum_{\beta_a,\beta_b}
 \left\{
    \begin{array}{ccc}
        {\bf R}'_1		& {\bf r}_a		& {\bf R}_1,\beta_a   \\
        {\bf R}'_2		& {\bf r}_b		& {\bf R}_2,\beta_b   \\
        {\bf R}',\gamma'	& {\bf r},\alpha	& {\bf R},\gamma
    \end{array}
\right\}_\beta
\langle {\bf R}_1||T_1^{{\bf r}_a}||{\bf R}'_1\rangle
\langle {\bf R}_2||T_2^{{\bf r}_b}||{\bf R}'_2\rangle.\nonumber
\end{eqnarray}
The $SU(3)$ 9j symbol can be calculated by evaluating
\begin{eqnarray}\label{EqAppC6}
&& \sum_\beta
 \left(
    \begin{array}{cc||c}
        {\bf R}' 	& {\bf r}	& {\bf R}  \\
        \{Y',I'\} 	& \{y,i\}	&\{Y,I\}
    \end{array}
\right)_\beta
 \left\{
    \begin{array}{ccc}
        {\bf R}'_1		& {\bf r}_a		& {\bf R}_1,\beta_a   \\
        {\bf R}'_2		& {\bf r}_b		& {\bf R}_2,\beta_b   \\
        {\bf R}',\gamma'	& {\bf r},\alpha	& {\bf R},\gamma
    \end{array}
\right\}_\beta \nonumber\\
&& = \sum_{\stackrel {\scriptstyle I_1, I'_1, I_2, I'_2, i_a, i_b,}{ Y_1, Y'_1, Y_2, Y'_2, y_a, y_b}}
f(I'_1, i_a, I_1; I'_2, i_b, I_2; I', i, I; I'_z, i_z)
 D({\bf R})
 \left(
    \begin{array}{cc||c}
        {\bf R}'_1  	& {\bf R}'_2	& {\bf R}'   \\
	\{Y'_1,I'_1\}	& \{Y'_2,I'_2\}	& \{Y',I'\}
    \end{array}
\right)_{\gamma'}\nonumber\\
&& \times
\left(
    \begin{array}{cc||c}
        {\bf r}_a 	& {\bf r}_b	& {\bf r}   \\
	\{y_a,i_a\}	& \{y_b,i_b\}	& \{y,i\}
    \end{array}
\right)_\alpha
\left(
    \begin{array}{cc||c}
        {\bf R}_1	& {\bf R}_2 	& {\bf R}   \\
	\{Y_1,I_1\}	& \{Y_2,I_2\}	& \{Y,I\}
    \end{array}
\right)_{\gamma}\nonumber\\
&& \times
 \left(
    \begin{array}{cc||c}
        {\bf R}'_1	& {\bf r}_a	& {\bf R}_1   \\
	\{Y'_1,I'_1\}	& \{y_a,i_a\}	& \{Y_1,I_1\}
    \end{array}
\right)_{\beta_a}
 \left(
    \begin{array}{cc||c}
        {\bf R}'_2  	& {\bf r}_b	& {\bf R}_2   \\
	\{Y'_2,I'_2\}	& \{y_b,i_b\}	& \{Y_2,I_2\}
    \end{array}
\right)_{\beta_b},
\end{eqnarray}
where the $SU(3)$ isoscalar factors are defined by
\begin{equation}
 \left(
    \begin{array}{cc|c}
        {\bf r}_a 		& {\bf r}_b		& {\bf r}   \\
	\{y_a,i_a,i_{az}\}	& \{y_b,i_b,i_{bz}\}	& \{y,i,i_z\}
    \end{array}
\right)_\alpha\
= \left(
    \begin{array}{cc|c}
        i_a 	& i_b 		& i   \\
	i_{az}	& i_{bz} 	& i_z
    \end{array}
\right)
\left(
    \begin{array}{cc||c}
        {\bf r}_a 	& {\bf r}_b 	& {\bf r}   \\
	\{y_a,i_a\}	& \{y_b,i_b\} 	& \{y,i\}
    \end{array}
\right)_\alpha.\nonumber
\end{equation}
The $f$ function in Eq.~(\ref{EqAppC6}) is given by
\begin{eqnarray}
 f(I'_1, i_a, I_1; I'_2, i_b, I_2; I', i, I; I'_z, i_z) =
 \sum_{I'_{1z}, i_{az}}
 \left(
    \begin{array}{cc|c}
        I'_1 	& I'_2 		& I'   \\
	I'_{1z}	& I'_z-I'_{1z} 	& I'_z
    \end{array}
\right)
 \left(
    \begin{array}{cc|c}
        i_a 	& i_b 		& i   \\
	i_{az}	& i_z-i_{az} 	& i_z
    \end{array}
\right)\nonumber\\
 \times\left(
    \begin{array}{cc|c}
        I_1 		& I_2 				& I   \\
	I'_{1z}+i_{az}	& I'_z-I'_{1z}+i_z-i_{az}	& I'_z+i_z
    \end{array}
\right)
 \left(
    \begin{array}{cc|c}
        I'_1 	& i_a 		& I_1   \\
	I'_{1z}	& i_{az} 	& I'_{1z} + i_{az}
    \end{array}
\right)\nonumber\\
 \times\left(
    \begin{array}{cc|c}
        I'_2 		& i_b 		& I_2   \\
	I'_z-I'_{1z}	& i_z-i_{az}	 	& I'_z-I'_{1z} + i_z-i_{az}
    \end{array}
\right)
 \left(
    \begin{array}{cc|c}
        I' 	& i 	& I   \\
	I'_z	& i_z	& I'_z + i_z
    \end{array}
\right)^{-1}.
\end{eqnarray}
The required explicit expression for the $SU(3)$ isoscalar factors have been obtained in Refs.~\cite{Hecht:1965,Vergados:1968sha}.

Expressions for $\langle S_c,{\bf R}_c ||\Lambda^{[s_b,{\bf r}_b]}|| S'_c,{\bf R}'_c \rangle_{\gamma_a}$ in Eq.~(\ref{EqAppC5}) are given explicitly for \mbox{$(N_c-1)$-quarks} symmetric cores in the following.

The matrix elements of the $G_{c}$ operator for symmetric cores are given by
\begin{eqnarray}\label{AppB5}
 \langle S_c, {\bf R}_{c} || G_c || S'_c, {\bf R}'_c \rangle_{\gamma_b}=
\left\{
    \begin{array}{cl}
        h_{\gamma_b} (S_c)		& \text{if } S_c=S'_c ,\\
        f (S_c,S'_c)			& \text{if } |S_c-S'_c|=1 \text{ and } {\gamma_b}=1 ,\\
        0							& \text{otherwise},
    \end{array}\right.
\end{eqnarray}
where
\begin{eqnarray}\label{AppB6}
 h_1&=&(-1)^{\delta_{2 S_c,N_c-1}}
 \frac{(2 S_c+1) (N_c+2)}{\sqrt{6} }
 \frac{ \sqrt{S_c (S_c+1) (N_c+1-2 S_c) (N_c+3+2 S_c)} }{\sqrt{12 S_c (S_c+1)+(N_c+5) (N_c-1)}},\nonumber\\
  h_2&=&-\frac{(1-\delta _{S_c,0})}{8 \sqrt{2}}
  \sqrt{\left(N_c+3\right)^2-4 S_c^2} \sqrt{N_c^2-1-4 S_c (S_c+1)}  \\
  &&\times \ (2 S_c+1)\sqrt{\frac{(N_c+1-2 S_c) (N_c+5+2 S_c)}{12 S_c (S_c+1)+(N_c+5) (N_c-1)}},	\nonumber\\
  f&=&-\frac{\sqrt{(2S'_c+1)(2S_c+1)}}{8\sqrt{2}} \sqrt{(N_c+1-2 S'_c) (N_c+3+2 S'_c) (N_c+1-2 S_c) (N_c+3+2 S_c)}\nonumber.
\end{eqnarray}

On another hand, for $T_c$ we have
\begin{eqnarray}
 \langle S_c, {\bf R}_{c} || T_c || S'_c, {\bf R}'_c \rangle=\delta_{S_c S'_c}{\hat S_c}\delta_{{\bf R}{\bf R'}}(-1)^{\delta_{q,0}}\sqrt{D({\bf R})C_{SU(3)}({\bf R})} \ ,
\end{eqnarray}
where $C_{SU(3)}({\bf R})$  is the quadratic Casimir operator for $SU(3)$ given by $C_{SU(3)}(p,q)=\frac{p^2+q^2+pq+3p+3q}{3}$.

The $S_c$ matrix elements for symmetric cores are
\begin{eqnarray}
 \langle S_c, {\bf R}_{c} || S_c || S'_c, {\bf R}'_c \rangle=\delta_{{\bf R}{\bf R'}}\sqrt{D({\bf R})}\delta_{S_c S'_c}\sqrt{S_c(S_c+1)(2S_c+1)} \ .
\end{eqnarray}

\section{Partial-wave amplitudes for symmetric and mixed-symmetric states}\label{AppB}

In this appendix we list the partial-wave amplitudes containing resonances with quantum numbers corresponding to S and MS states of the ${\cal N}=2$ band in large the $N_c$ quark picture.

\begin{table}[H]
\centering
\begin{tabular}{ccl}
\hline\hline
 State		& Pole mass	& Partial wave, $K$-amplitudes					\\
 \hline
 $N_{1/2}$	& $m_0$, $m_{1^\pm}$		& $P_{11}^{\pi N}$=$\frac13 \left(\tau_{01}^{\pi} + 2 \tau_{11}^{\pi}\right)$		\\
		& 				& $P_{11}^{\eta N}$=$\tau_{11}^{\eta}$							\\
 $N_{3/2}$	& $m_{1^\pm}$, $m_{2^\pm}$	& $P_{13}^{\pi N}$=$\frac16 \left(\tau_{11}^{\pi} + 5 \tau_{21}^{\pi}\right)$		\\
		& 				& $P_{13}^{\eta N}$=$\tau_{11}^{\eta}$						\\
 $N_{5/2}$	& $m_{2^\pm}$, $m_3$	& $F_{15}^{\pi N}$=$\frac19 \left(5 \tau_{23}^{\pi} + 4 \tau_{33}^{\pi}\right)$		\\
		& 			& $F_{15}^{\eta N}$=$\tau_{33}^{\eta}$		\\
 $N_{7/2}$	& $m_3$			& $F_{17}^{\pi N}$=$\frac14 \left(\tau_{33}^{\pi} + 3 \tau_{43}^{\pi}\right)$		\\
		& 			& $F_{17}^{\eta N}$=$\tau_{33}^{\eta}$		\\
 \hline
 $\Delta_{1/2}$	& $m_{1^\pm}$, $m_{2^\pm}$	& $P_{31}^{\pi N}$=$\frac16 \left(\tau_{11}^{\pi} + 5 \tau_{21}^{\pi}\right)$		\\
 $\Delta_{3/2}$	& $m_0$, $m_{1^\pm}$, $m_{2^\pm}$, $m_3$	& $P_{33}^{\pi N}$=$\frac{1}{12} \left(2 \tau_{01}^{\pi} + 5 \tau_{11}^{\pi} + 5 \tau_{21}^{\pi}\right)$		\\
		& 						& $P_{33}^{\pi \Delta}$=$\frac{1}{15} \left(5 \tau_{01}^{\pi} + 2 \tau_{11}^{\pi} + 8 \tau_{21}^{\pi}\right)$		\\
		& 						& $F_{33}^{\pi \Delta}$=$\frac{1}{5} \left(\tau_{23}^{\pi} + 4 \tau_{33}^{\pi}\right)$		\\
		& 						& $P_{33}^{\eta \Delta}$=$\tau_{11}^{\eta}$		\\
		& 						& $F_{33}^{\eta \Delta}$=$\tau_{33}^{\eta}$		\\
 $\Delta_{5/2}$	& $m_{1^\pm}$, $m_{2^\pm}$, $m_3$	& $P_{35}^{\pi N}$=$\frac{1}{126} \left(10 \tau_{23}^{\pi} + 35 \tau_{33}^{\pi} + 81 \tau_{43}^{\pi}\right)$		\\
		& 					& $P_{35}^{\pi \Delta}$=$\frac{1}{10} \left(3 \tau_{11}^{\pi} + 7 \tau_{21}^{\pi}\right)$		\\
		& 					& $F_{35}^{\pi \Delta}$=$\frac{1}{1260} \left(512 \tau_{23}^{\pi} + 343 \tau_{33}^{\pi} + 405 \tau_{43}^{\pi}\right)$		\\
		& 					& $P_{35}^{\eta \Delta}$=$\tau_{11}^{\eta}$		\\
		& 					& $F_{35}^{\eta \Delta}$=$\tau_{33}^{\eta}$		\\
 $\Delta_{7/2}$	& $m_{2^\pm}$, $m_3$	& $F_{37}^{\pi N}$=$\frac{1}{56} \left(20 \tau_{23}^{\pi} + 21 \tau_{33}^{\pi} + 15 \tau_{43}^{\pi}\right)$		\\
\hline\hline
\end{tabular}\caption{Large $N_c$ mass eigenvalues in the $1/N_c$ expansion corresponding to states in the $[{\bf 56},L^+]$ and $[{\bf 70},L^+]$ multiplets with $L=0,2$ and the partial-wave amplitudes containing resonances with the same quantum numbers. Since $I=I'$ in all the partial-waves, to lighten notation we replaced the $II'Y$ labels in $\tau$ to those of a meson having the corresponding quantum numbers. Partial-wave amplitudes containing the nonstrange resonances were calculated before in Ref.~\cite{Cohen:2003nn} and are listed here for the reader's convenience.}
\label{TabAppB1}
\end{table}

\begin{table}[H]
\begin{tabular}{ccl}
\hline\hline
 State		& Pole mass	& Partial wave, $K$-amplitudes					\\
\hline
 $\Lambda^{\bf 8}_{1/2}$	& $m_0$, $m_{1^\pm}$	& $P_{01}^{\pi \Sigma}$=$\frac13 \left(\tau_{01}^{\pi} + 2 \tau_{11}^{\pi}\right)$		\\
				&	 		& $P_{01}^{\eta \Lambda}$=$\tau_{11}^{\eta}$		\\
 $\Lambda^{\bf 8}_{3/2}$	& $m_{1^\pm}$, $m_{2^\pm}$	& $P_{03}^{\pi \Sigma}$=$\frac16 \left(\tau_{11}^{\pi} + 5 \tau_{21}^{\pi}\right)$		\\
				&	 			& $P_{03}^{\eta \Lambda}$=$\tau_{11}^{\eta}$		\\
 $\Lambda^{\bf 8}_{5/2}$	& $m_{2^\pm}$, $m_3$	& $F_{05}^{\pi \Sigma}$=$\frac19 \left(5\tau_{23}^{\pi} + 4 \tau_{33}^{\pi}\right)$		\\
				&	 		& $F_{05}^{\eta \Lambda}$=$\tau_{33}^{\eta}$		\\
 $\Lambda^{\bf 8}_{7/2}$	& $m_3$	& $F_{07}^{\pi \Sigma}$=$\frac14 \left(\tau_{33}^{\pi} + 3 \tau_{43}^{\pi}\right)$		\\
				&	& $F_{07}^{\eta \Lambda}$=$\tau_{33}^{\eta}$		\\
				\hline
 $\Lambda^{\bf 1}_{1/2}$	& $m_\frac12$	& $P_{01}^{{\bar K} N}$=$\tau_{\frac121}^{\bar K}$		\\
 $\Lambda^{\bf 1}_{3/2}$	& $m_\frac32$	& $P_{03}^{{\bar K} N}$=$\tau_{\frac321}^{\bar K}$		\\
 $\Lambda^{\bf 1}_{5/2}$	& $m_\frac52$	& $F_{05}^{{\bar K} N}$=$\tau_{\frac523}^{\bar K}$		\\
\vspace{-5pt}\\
\hline\hline
\end{tabular}
\quad
\begin{tabular}{ccl}
\hline\hline
 State		& Pole mass	& Partial wave, $K$-amplitudes					\\
\hline
 $\Sigma^{\bf 8}_{1/2}$		& $m_0$, $m_{1^\pm}$	& $P_{11}^{\pi \Lambda}$=$\frac13 \left(\tau_{01}^{\pi} + 2 \tau_{11}^{\pi}\right)$		\\
				& 	 		& $P_{11}^{\pi \Sigma}$=$\frac13 \left(\tau_{01}^{\pi} + 2 \tau_{11}^{\pi}\right)$		\\
				& 	 		& $P_{11}^{\eta \Lambda}$=$\tau_{11}^{\eta}$		\\
 $\Sigma^{\bf 8}_{3/2}$		& $m_{1^\pm}$, $m_{2^\pm}$	& $P_{13}^{\pi \Lambda}$=$\frac16 \left(\tau_{11}^{\pi} + 5 \tau_{21}^{\pi}\right)$		\\
				& 				& $P_{13}^{\pi \Sigma}$=$\frac16 \left(\tau_{11}^{\pi} + 5 \tau_{21}^{\pi}\right)$		\\
				& 				& $P_{13}^{\eta \Lambda}$=$\tau_{11}^{\eta}$		\\
 $\Sigma^{\bf 8}_{5/2}$		& $m_{2^\pm}$, $m_3$	& $F_{15}^{\pi \Lambda}$=$\frac19 \left(5 \tau_{23}^{\pi} + 4 \tau_{33}^{\pi}\right)$		\\
				& 			& $F_{15}^{\pi \Sigma}$=$\frac19 \left(4 \tau_{23}^{\pi} + 5 \tau_{33}^{\pi}\right)$		\\
				& 			& $F_{15}^{\eta \Sigma}$=$\tau_{33}^{\eta}$		\\
 $\Sigma^{\bf 8}_{7/2}$	& $m_3$	& $F_{17}^{\pi \Lambda}$=$\frac14 \left(\tau_{33}^{\pi} + 3 \tau_{43}^{\pi}\right)$		\\
				& 	& $F_{17}^{\pi \Sigma}$=$\frac14 \left(\tau_{33}^{\pi} + 3 \tau_{43}^{\pi}\right)$		\\
				& 	& $F_{17}^{\eta \Sigma}$=$\tau_{33}^{\eta}$		\\
\hline\hline
\end{tabular}\caption{Continuation of Tab.~\ref{TabAppB1}}
\label{TabAppB3}
\end{table}

\begin{table}[H]
\centering
\begin{tabular}{ccl}
\hline\hline
 State		& Pole mass	& Partial wave, $K$-amplitudes					\\
\hline
 $\Sigma^{\bf 10}_{1/2}$	& $m_{1^\pm}$, $m_{2^\pm}$	& $P_{11}^{\pi \Lambda}$=$\frac16 \left(\tau_{11}^{\pi} + 5 \tau_{21}^{\pi}\right)$		\\
				& 	 			& $P_{11}^{\pi \Sigma}$=$\frac16 \left(\tau_{11}^{\pi} + 5 \tau_{21}^{\pi}\right)$		\\
 $\Sigma^{\bf 10}_{3/2}$	& $m_0$, $m_{1^\pm}$, $m_{2^\pm}$, $m_3$	& $P_{13}^{\pi \Lambda}$=$\frac{1}{12} \left(2 \tau_{01}^{\pi} + 5 \tau_{11}^{\pi} + 5 \tau_{21}^{\pi}\right)$		\\
				& 	 					& $P_{13}^{\pi \Sigma}$=$\frac{1}{12} \left(2 \tau_{01}^{\pi} + 5 \tau_{11}^{\pi} + 5 \tau_{21}^{\pi}\right)$		\\
				& 			 			& $P_{13}^{\pi \Sigma^*}$=$\frac{1}{15} \left(5 \tau_{01}^{\pi} + 2 \tau_{11}^{\pi} + 8 \tau_{21}^{\pi}\right)$		\\
				& 			 			& $F_{13}^{\pi \Sigma^*}$=$\frac15 \left(\tau_{23}^{\pi} + 4 \tau_{33}^{\pi}\right)$		\\
				& 	 					& $P_{13}^{\eta \Sigma^*}$=$\tau_{11}^{\eta}$		\\
				& 	 					& $F_{13}^{\eta \Sigma^*}$=$\tau_{33}^{\eta}$		\\
 $\Sigma^{\bf 10}_{5/2}$	& $m_{1^\pm}$, $m_{2^\pm}$, $m_3$	& $F_{15}^{\pi \Lambda}$=$\frac{1}{126} \left(10 \tau_{23}^{\pi} + 35 \tau_{33}^{\pi} + 81 \tau_{43}^{\pi}\right)$		\\
				& 					& $F_{15}^{\pi \Sigma}$=$\frac{1}{126} \left(10 \tau_{23}^{\pi} + 35 \tau_{33}^{\pi} + 81 \tau_{43}^{\pi}\right)$		\\
				& 		 			& $P_{15}^{\pi \Sigma^*}$=$\frac{1}{10} \left(3 \tau_{11}^{\pi} + 7 \tau_{21}^{\pi}\right)$		\\
				& 		 			& $F_{15}^{\pi \Sigma^*}$=$\frac{1}{1260} \left(512 \tau_{23}^{\pi} + 343 \tau_{33}^{\pi} + 405 \tau_{43}^{\pi}\right)$		\\
				& 					& $P_{15}^{\eta \Sigma^*}$=$\tau_{11}^{\eta}$		\\
				& 					& $F_{15}^{\eta \Sigma^*}$=$\tau_{33}^{\eta}$		\\
 $\Sigma^{\bf 10}_{7/2}$	& $m_{2^\pm}$, $m_3$	& $F_{15}^{\pi \Lambda}$=$\frac{1}{126} \left(10 \tau_{23}^{\pi} + 35 \tau_{33}^{\pi} + 81 \tau_{43}^{\pi}\right)$		\\
				& 			& $F_{15}^{\pi \Sigma}$=$\frac{1}{126} \left(10 \tau_{23}^{\pi} + 35 \tau_{33}^{\pi} + 81 \tau_{43}^{\pi}\right)$		\\
\hline
 $\Sigma^{\bf S}_{1/2}$		& $m_\frac12$, $m_\frac32$		& $P_{11}^{{\bar K} N}$=$\frac19 \left(\tau_{\frac121}^{\bar K} + 8 \tau_{\frac321}^{\bar K}\right)$		\\
 $\Sigma^{\bf S}_{3/2}$		& $m_\frac12$, $m_\frac32$, $m_\frac52$	& $P_{13}^{{\bar K} N}$=$\frac19 \left(4 \tau_{\frac121}^{\bar K} + 5 \tau_{\frac321}^{\bar K}\right)$		\\
				& 					& $P_{13}^{{\bar K} \Delta}$=$\frac19 \left(5 \tau_{\frac121}^{\bar K} + 4 \tau_{\frac321}^{\bar K}\right)$		\\
				& 					& $F_{13}^{{\bar K} \Delta}$=$\tau_{\frac523}^{\bar K}$		\\
 $\Sigma^{\bf S}_{5/2}$		& $m_\frac32$, $m_\frac52$	& $F_{15}^{{\bar K} N}$=$\frac{1}{21} \left(5 \tau_{\frac523}^{\bar K} + 16 \tau_{\frac723}^{\bar K}\right)$		\\
				& 				& $P_{15}^{{\bar K} \Delta}$=$\tau_{\frac321}^{\bar K}$		\\
				& 				& $F_{15}^{{\bar K} \Delta}$=$\frac{1}{21} \left(16 \tau_{\frac523}^{\bar K} + 5 \tau_{\frac723}^{\bar K}\right)$		\\
 $\Sigma^{\bf S}_{7/2}$		& $m_\frac52$			& $F_{17}^{{\bar K} N}$=$\frac17 \left(4 \tau_{\frac523}^{\bar K} + 3 \tau_{\frac723}^{\bar K}\right)$		\\
\hline\hline
\end{tabular}\caption{Continuation of Tab.~\ref{TabAppB3}}
\label{TabAppB4}
\end{table}

\begin{table}[H]
\centering
\begin{tabular}{ccl}
\hline\hline
 State		& Pole mass	& Partial wave, $K$-amplitudes					\\
 \hline
 $\Xi^{\bf 8}_{1/2}$	& $m_0$, $m_{1^\pm}$	& $P_{11}^{\pi \Xi}$=$\frac{1}{27} \left(\tau_{01}^{\pi} + 2 \tau_{11}^{\pi}\right)$		\\
			& 			& $P_{11}^{\eta \Xi}$=$\tau_{11}^{\eta}$		\\
 $\Xi^{\bf 8}_{3/2}$	& $m_{1^\pm}$, $m_{2^\pm}$	& $P_{13}^{\pi \Xi}$=$\frac{1}{54} \left(\tau_{11}^{\pi} + 5 \tau_{21}^{\pi}\right)$		\\
			& 				& $P_{13}^{\eta \Xi}$=$\tau_{11}^{\eta}$		\\
 $\Xi^{\bf 8}_{5/2}$	& $m_{2^\pm}$, $m_3$	& $F_{15}^{\pi \Xi}$=$\frac{1}{81} \left(5\tau_{23}^{\pi} + 4 \tau_{33}^{\pi}\right)$		\\
			& 			& $F_{15}^{\eta \Xi}$=$\tau_{33}^{\eta}$		\\
 $\Xi^{\bf 8}_{7/2}$	& $m_3$			& $F_{17}^{\pi \Xi}$=$\frac{1}{36} \left(\tau_{33}^{\pi} + 3 \tau_{43}^{\pi}\right)$		\\
			& 			& $F_{17}^{\eta \Xi}$=$\tau_{33}^{\eta}$		\\
 $\Xi^{\bf 10}_{1/2}$	& $m_{1^\pm}$, $m_{2^\pm}$			& $P_{11}^{\pi \Xi}$=$\frac{4}{27} \left(\tau_{11}^{\pi} + 5 \tau_{21}^{\pi}\right)$		\\
 $\Xi^{\bf 10}_{3/2}$	& $m_0$, $m_{1^\pm}$, $m_{2^\pm}$, $m_3$	& $P_{13}^{\pi \Xi}$=$\frac{2}{27} \left(2 \tau_{01}^{\pi} + 5 \tau_{11}^{\pi} + 5 \tau_{21}^{\pi} \right)$	\\
			& 						& $P_{13}^{\pi \Xi^*}$=$\frac{5}{108} \left(2 \tau_{01}^{\pi} + 5 \tau_{11}^{\pi} + 5 \tau_{21}^{\pi} \right)$	\\
			& 						& $P_{13}^{\eta \Xi^*}$=$\tau_{11}^{\eta}$	\\
			& 						& $F_{13}^{\eta \Xi^*}$=$\tau_{33}^{\eta}$	\\
 $\Xi^{\bf 10}_{5/2}$	& $m_{1^\pm}$, $m_{2^\pm}$, $m_3$	& $F_{15}^{\pi \Xi}$=$\frac{8}{1134} \left(10 \tau_{23}^{\pi} + 35 \tau_{33}^{\pi} + 81 \tau_{43}^{\pi} \right)$	\\
			& 					& $F_{15}^{\pi \Xi^*}$=$\frac{5}{1134} \left(10 \tau_{23}^{\pi} + 35 \tau_{33}^{\pi} + 81 \tau_{43}^{\pi} \right)$	\\
			& 					& $P_{15}^{\eta \Xi^*}$=$\tau_{11}^{\eta}$	\\
			& 					& $F_{15}^{\eta \Xi^*}$=$\tau_{33}^{\eta}$	\\
 $\Xi^{\bf 10}_{7/2}$	& $m_{2^\pm}$, $m_3$	& $F_{17}^{\pi \Xi}$=$\frac{1}{63} \left(20 \tau_{23}^{\pi} + 21 \tau_{33}^{\pi} + 15 \tau_{43}^{\pi} \right)$	\\
\hline\hline
\end{tabular}\caption{Continuation of Tab.~\ref{TabAppB4}}
\label{TabAppB5}
\end{table}

\begin{table}[H]
\centering
\begin{tabular}{ccl}
\hline\hline
 State		& Pole mass	& Partial wave, $K$-amplitudes					\\
 \hline
 $\Xi^{\bf 1}_{1/2}$	& $m_\frac12$	& $P_{11}^{\bar K \Sigma}$=$\tau_{\frac121}^{\bar K}$		\\
			& 		& $P_{11}^{\bar K \Lambda}$=$\tau_{\frac121}^{\bar K}$		\\
 $\Xi^{\bf 1}_{3/2}$	& $m_\frac32$	& $P_{13}^{\bar K \Sigma}$=$\tau_{\frac321}^{\bar K}$		\\
			& 		& $P_{13}^{\bar K \Lambda}$=$\tau_{\frac321}^{\bar K}$		\\
 $\Xi^{\bf 1}_{5/2}$	& $m_\frac52$	& $F_{11}^{\bar K \Sigma}$=$\tau_{\frac523}^{\bar K}$		\\
			& 		& $F_{11}^{\bar K \Sigma}$=$\tau_{\frac523}^{\bar K}$		\\
 $\Xi^{\bf S}_{1/2}$	& $m_\frac12$, $m_\frac32$		& $P_{11}^{\bar K \Sigma}$=$\frac19 \left( \tau_{\frac121}^{\bar K} + 8 \tau_{\frac321}^{\bar K} \right)$	\\
			&					& $P_{11}^{\bar K \Lambda}$=$\frac19 \left( \tau_{\frac121}^{\bar K} + 8 \tau_{\frac321}^{\bar K} \right)$		\\
 $\Xi^{\bf S}_{3/2}$	& $m_\frac12$, $m_\frac32$, $m_\frac52$	& $P_{13}^{\bar K \Sigma}$=$\frac19 \left(4 \tau_{\frac121}^{\bar K} + 5 \tau_{\frac321}^{\bar K} \right)$		\\
			& 					& $P_{13}^{\bar K \Lambda}$=$\frac19 \left(4 \tau_{\frac121}^{\bar K} + 5 \tau_{\frac321}^{\bar K} \right)$		\\
			& 					& $P_{13}^{\bar K \Sigma^*}$=$\frac19 \left(5 \tau_{\frac121}^{\bar K} + 4 \tau_{\frac321}^{\bar K} \right)$		\\
			& 					& $F_{13}^{\bar K \Sigma^*}$=$\tau_{\frac523}^{\bar K}$									\\
 $\Xi^{\bf S}_{5/2}$	& $m_\frac32$, $m_\frac52$		& $F_{15}^{\bar K \Sigma}$=$\frac{1}{21} \left(5 \tau_{\frac523}^{\bar K} + 16 \tau_{\frac723}^{\bar K} \right)$		\\
			& 					& $F_{15}^{\bar K \Lambda}$=$\frac{1}{21} \left(5 \tau_{\frac523}^{\bar K} + 16 \tau_{\frac723}^{\bar K} \right)$		\\
			& 					& $P_{15}^{\bar K \Sigma^*}$=$\tau_{\frac321}^{\bar K}$									\\
			& 					& $F_{15}^{\bar K \Sigma^*}$=$\frac{1}{21} \left(16 \tau_{\frac523}^{\bar K} + 5 \tau_{\frac723}^{\bar K} \right)$		\\
 $\Xi^{\bf S}_{7/2}$	& $m_\frac52$				& $F_{17}^{\bar K \Sigma}$=$\frac17 \left(4 \tau_{\frac523}^{\bar K} + 3 \tau_{\frac723}^{\bar K} \right)$		\\
			& 					& $F_{17}^{\bar K \Lambda}$=$\frac17 \left(4 \tau_{\frac523}^{\bar K} + 3 \tau_{\frac723}^{\bar K} \right)$		\\
\hline\hline
\end{tabular}\caption{Continuation of Tab.~\ref{TabAppB5}}
\label{TabAppB6}
\end{table}

\begin{table}[H]
\centering
\begin{tabular}{ccl}
\hline\hline
 State		& Pole mass	& Partial wave, $K$-amplitudes					\\
\hline
 $\Omega^{\bf 10}_{1/2}$	& $m_{1^\pm}$, $m_{2^\pm}$			& $P_{01}^{\eta \Omega}$=$\tau_{11}^{\eta}$						\\
				& 	 					& $P_{01}^{\pi \Omega'}$=$\frac16 \left(5 \tau_{11}^{\pi} + \tau_{21}^{\pi}\right)$	\\
 $\Omega^{\bf 10}_{3/2}$	& $m_0$, $m_{1^\pm}$, $m_{2^\pm}$, $m_3$	& $P_{03}^{\eta \Omega}$=$\tau_{11}^{\eta}$						\\
				& 						& $F_{03}^{\eta \Omega}$=$\tau_{33}^{\eta}$						\\
				& 	 					& $P_{03}^{\pi \Omega'}$=$\frac{1}{15} \left(5 \tau_{01}^{\pi} + 2 \tau_{11}^{\pi} + 8 \tau_{21}^{\pi}\right)$	\\
				& 	 					& $F_{03}^{\pi \Omega'}$=$\frac15 \left(\tau_{23}^{\pi} + 4 \tau_{33}^{\pi}\right)$	\\
 $\Omega^{\bf 10}_{5/2}$	& $m_{1^\pm}$, $m_{2^\pm}$, $m_3$		& $P_{05}^{\eta \Omega}$=$\tau_{11}^{\eta}$						\\
				& 						& $F_{05}^{\eta \Omega}$=$\tau_{33}^{\eta}$						\\
				& 	 					& $P_{05}^{\pi \Omega'}$=$\frac{1}{10} \left(3 \tau_{11}^{\pi} + 7 \tau_{21}^{\pi} \right)$	\\
				& 	 					& $F_{05}^{\pi \Omega'}$=$\frac{128}{135} \tau_{23}^{\pi} + \frac{49}{180} \tau_{33}^{\pi} + \frac{9}{28} \tau_{43}^{\pi}$	\\
 $\Omega^{\bf 10}_{7/2}$	& $m_{2^\pm}$, $m_3$				& $F_{07}^{\eta \Omega}$=$\tau_{33}^{\eta}$						\\
				& 						& $H_{07}^{\eta \Omega}$=$\tau_{55}^{\eta}$						\\
				& 	 					& $F_{07}^{\pi \Omega'}$=$\frac{1}{7} \left(3 \tau_{23}^{\pi} + 4 \tau_{43}^{\pi} \right)$	\\
				& 	 					& $H_{07}^{\pi \Omega'}$=$\frac{1}{25} \left(7 \tau_{45}^{\pi} + 18 \tau_{55}^{\pi} \right)$	\\
\hline
 $\Omega^{\bf S}_{1/2}$		& $m_\frac12$, $m_\frac32$		& $P_{01}^{{\bar K} \Xi}$=$\frac19 \left(\tau_{\frac121}^{\bar K} + 8 \tau_{\frac321}^{\bar K}\right)$		\\
 $\Omega^{\bf S}_{3/2}$		& $m_\frac12$, $m_\frac32$, $m_\frac52$	& $P_{03}^{{\bar K} \Xi}$=$\frac19 \left(4 \tau_{\frac121}^{\bar K} + 5 \tau_{\frac321}^{\bar K}\right)$		\\
				& 					& $P_{03}^{{\bar K} \Xi^*}$=$\frac19 \left(5 \tau_{\frac121}^{\bar K} + 4 \tau_{\frac321}^{\bar K}\right)$		\\
				& 					& $F_{03}^{{\bar K} \Xi^*}$=$ \tau_{\frac523}^{\bar K}$		\\
 $\Omega^{\bf S}_{5/2}$		& $m_\frac32$, $m_\frac52$		& $F_{05}^{{\bar K} \Xi}$=$\frac{1}{25} \left(5 \tau_{\frac523}^{\bar K} + 16 \tau_{\frac723}^{\bar K}\right)$		\\
				& 					& $P_{05}^{{\bar K} \Xi^*}$=$ \tau_{\frac321}^{\bar K}$		\\
				& 					& $F_{05}^{{\bar K} \Xi^*}$=$\frac{1}{21} \left(16 \tau_{\frac523}^{\bar K} + 5 \tau_{\frac723}^{\bar K}\right)$		\\
 $\Omega^{\bf S}_{7/2}$		& $m_\frac52$				& $F_{07}^{{\bar K} \Xi}$=$\frac17 \left(4 \tau_{\frac523}^{\bar K} + 3 \tau_{\frac723}^{\bar K}\right)$		\\
\hline\hline
\end{tabular}\caption{Continuation of Tab.~\ref{TabAppB6}}
\label{TabAppB7}
\end{table}

\section{Core composition of antisymmetric states}\label{AppE}
To find the core composition of the antisymmetric states of the ${\cal N}=2$ band, we used the quadratic Casimir operator which can be defined as $C_{R}=\sum_a\Lambda_a\Lambda_a$ where $\Lambda_a$ are the generators of the representation R. When working with $SU(6)$, it is useful to recall the relation $C_{SU(6)}=2 G_{ia}G_{ia}+\frac12C_{SU(3)}+\frac13C_{SU(2)}$ \cite{Goity:2002pu}.
The $SU(6)$ generators can be expressed in terms of core and quarks operators \mbox{$S_i=s_i+(S_c)_i$}, \mbox{$T_a=t_a+(T_c)_a$}, \mbox{$G_{ia}=g_{ia}+(G_c)_{ia}$}, then
\begin{equation} \label{casimir}
 C_{SU(6)}=C_{SU(6)}^c - \frac12 C_{SU(3)}^c - \frac13 C_{SU(2)}^c + 4 g_{ia}(G_c)_{ia} + 2 g_{ia}g_{ia} + \frac12 C_{SU(3)} + \frac13 C_{SU(2)},
\end{equation}
where $g_{ia}g_{ia}=s^2t^2$ so that $\langle g_{ia}g_{ia}\rangle=1$ and $C_{R}^c$ denote a quadratic Casimir operator built with core generators.
The Casimir matrix element of a given representation can be expressed in terms of the boxes in its Young tableau (see Ref.~\cite{Gross:1993hu}), in particular for the A representation of $SU(6)$ whose multiplet in Dynkin notation is $(N_c-3,0,1,0,0)$ is given by
\begin{equation} \label{casimir2}
 C_{SU(6)}(N_c-3,0,1,0,0)=\frac{5N_c^2+6N_c}{12}.
\end{equation}
Then, when calculating the matrix elements of the operator in Eq.~(\ref{casimir}) all terms in the LHS and RHS are determined except for $\langle g_{ia}(G_c)_{ia} \rangle$.
As mentioned before, the core can be assumed to be an $(N_c-2)$ quarks core $\tilde c$ in a symmetric representation coupled to a quark so that the overall symmetry is mixed-symmetric. We can write $(G_c)_{ia}=(G_{\tilde c})_{ia}+g_{ia}$ and using the Wigner-Eckart theorems for $SU(2)$ and $SU(3)$ we find that
\begin{eqnarray}\label{Gcore}
{}^A\langle S, {\bf R}| g_{ia}(G_c)_{ia}|S',{\bf R}'\rangle^A &=& \sum_{i,i'}c_ic_{i'} \delta_{{\bf R},{\bf R}'}\delta_{S,S'} \frac{12\hat S}{D({\bf R})\sqrt{3 D({\bf R}_{c_{i}})}}
\left\{
    \begin{array}{ccc}
        S_{c_i} & S_{c_{i'}} 	& 1   \\
        1/2 	& 1/2		& 1   \\
        S 	& S'		& 0
    \end{array}
\right\} \nonumber\\
& \times &\sum_{\gamma_a}\left\{
    \begin{array}{ccc}
        {\bf R}_{c_i'}	& {\bf 8}	& {\bf R}_{c_i},\gamma_a   \\
        {\bf 3}		& {\bf 8}	& {\bf 3}   \\
        {\bf R}'	& {\bf 1}	& {\bf R}
    \end{array}
\right\}
\langle S_{c_{i}},{\bf R}_{c_{i}} || G_c^{[1,{\bf 8}]} || S_{c_{i'}},{\bf R}_{c_{i'}} \rangle_{\gamma_a},
\end{eqnarray}
where we used the relation between Cartesian and spherical basis given by \mbox{$g_{ia}(G_c)_{ia}=-\sqrt3\sqrt8\left(g_{ia}(G_c)_{ia}\right)^{[0,{\bf 1}]}_{0,0}$}. The second term in braces is an $SU(3)$ 9j symbol whose definition can be found in App.~\ref{AppC}. Details on how to calculate $\langle S_{c_{i}},{\bf R}_{c_{i}} || G_c^{[1,{\bf 8}]} || S_{c_{i'}},{\bf R}_{c_{i'}} \rangle_{\gamma_a}$ can be found on App.~\ref{AppD}.

Three distinct symmetries can result from coupling one quark to a mixed-symmetric core which can be written in Young diagrams as
\begin{equation}
\raisebox{-8.0pt}{\drawsquare{10.0}{0.4}}\hskip-10.4pt
        \raisebox{2pt}{\drawsquare{10.0}{0.4}}\hskip-0.4pt
        \raisebox{2pt}{\drawsquare{10.0}{0.4}}\hskip-0.4pt
        \raisebox{2pt}{\drawsquare{10.0}{0.4}}\hskip-0.4pt
        \raisebox{6pt}{ \ldots}
        \raisebox{2pt}{\drawsquare{10.0}{0.4}}\hskip-0.4pt
        \hskip4pt
        \times
        \hskip4pt
        \raisebox{-0.0pt}{\drawsquare{10.0}{0.4}}
        \hskip4pt
        =
        \hskip4pt
        \underbrace{\raisebox{-8.0pt}{\drawsquare{10.0}{0.4}}\hskip-10.4pt
	\raisebox{-18.0pt}{\drawsquare{10.0}{0.4}}\hskip-10.4pt
        \raisebox{2pt}{\drawsquare{10.0}{0.4}}\hskip-0.4pt
        \raisebox{2pt}{\drawsquare{10.0}{0.4}}\hskip-0.4pt
        \raisebox{2pt}{\drawsquare{10.0}{0.4}}\hskip-0.4pt
        \raisebox{6pt}{ \ldots}
        \raisebox{2pt}{\drawsquare{10.0}{0.4}}\hskip-0.4pt}_{A}
        \hskip4pt
        +
        \hskip4pt
        \underbrace{
        \raisebox{-8.0pt}{\drawsquare{10.0}{0.4}}\hskip-10.4pt
        \raisebox{2pt}{\drawsquare{10.0}{0.4}}\hskip-0.4pt
        \raisebox{2pt}{\drawsquare{10.0}{0.4}}\hskip-0.4pt
        \raisebox{2pt}{\drawsquare{10.0}{0.4}}\hskip-0.4pt
        \raisebox{6pt}{ \ldots}
        \raisebox{2pt}{\drawsquare{10.0}{0.4}}\hskip-0.4pt
        \raisebox{2pt}{\drawsquare{10.0}{0.4}}\hskip-0.4pt
        \raisebox{-18.0pt}{\drawsquare{0.0}{0.0}}\hskip-10.4pt
        \hskip16pt}_{MS}
        +
        \hskip4pt
        \underbrace{\raisebox{-8.0pt}{\drawsquare{10.0}{0.4}}\hskip-10.4pt
        \raisebox{2pt}{\drawsquare{10.0}{0.4}}\hskip-0.4pt
        \raisebox{2pt}{\drawsquare{10.0}{0.4}}\hskip-0.4pt
        \raisebox{2pt}{\drawsquare{10.0}{0.4}}\hskip-20.4pt
        \raisebox{-8.0pt}{\raisebox{0pt}{\drawsquare{10.0}{0.4}}\hskip-0.4pt}\hskip10.4pt
        \raisebox{6pt}{ \ldots}
        \raisebox{2pt}{\drawsquare{10.0}{0.4}}\hskip-0.4pt
        \raisebox{-18.0pt}{\drawsquare{0.0}{0.0}}\hskip-10.4pt
        \hskip16pt}_{MS2}
        \,,
\end{equation}
where the representations in Dynkin notation are \mbox{$A=(N_c-2,1,1,0,0)$}, \mbox{$MS=(N_c-1,1,0,0,0)$} and \mbox{$MS2=(N_c-2,2,0,0,0)$}. (Note that $MS2$ configuration is not possible for \mbox{$N_c=3$}.)
The Casimir invariants for the MS and MS2 representation are $C_{SU(6)}(N_c-2,2,0,0,0)=\frac{1}{12} N_c (5 N_c+6)+2$ and $ C_{SU(6)}(N_c-1,1,0,0,0,0)=\frac{1}{12} N_c (5 N_c+18)$ so that, when we diagonalize the matrices we obtain the core composition of the A, MS and MS2 representations. In Eq.~(\ref{statesanti3}) of Sec.~\ref{Sec7} we list the results for the A states which we used in the calculations that follow in that section.

\section{Partial-wave amplitudes for antisymmetric states}\label{AppB2}

In this Appendix we list the partial-wave amplitudes containing resonances with quantum numbers corresponding to A states of the ${\cal N}=2$ band in large the $N_c$ quark picture.

\begin{table}[H]
\begin{tabular}{ccl}
\hline\hline
 State		& Pole mass	& Partial wave, $K$-amplitudes				\\
 \hline
 $N_{1/2}$		& $m_0$, $m_1$	& $S_{11}^{\pi N}$=$\tau_{10}^{\pi}$		\\
			& 		& $S_{11}^{\eta N}$=$\tau_{00}^{\eta}$		\\
 $N_{3/2}$		& $m_1$, $m_2$	& $D_{13}^{\pi N}$=$\frac12 \left(\tau_{12}^{\pi}+ \tau_{22}^{\pi}\right)$		\\
			& 		& $D_{13}^{\eta N}$=$\tau_{22}^{\eta}$							\\
 $N_{5/2}$		& $m_2$ 	& $D_{13}^{\pi N}$=$\frac19 \left(2 \tau_{22}^{\pi}+ 7 \tau_{32}^{\pi}\right)$		\\
			& 		& $D_{13}^{\eta N}$=$\tau_{22}^{\eta}$							\\
													\\
													\\
													\\
 \hline
 $\Lambda^{\bf8}_{1/2}$	& $m_0$, $m_1$	& $S_{01}^{\pi \Sigma}$=$\tau_{10}^{\pi}$		\\
			& 		& $S_{01}^{\eta \Lambda}$=$\tau_{00}^{\eta}$		\\
 $\Lambda^{\bf8}_{3/2}$	& $m_1$, $m_2$	& $D_{03}^{\pi \Sigma}$=$\frac12 \left(\tau_{12}^{\pi}+\tau_{22}^{\pi}\right)$		\\
			& 		& $D_{03}^{\eta \Lambda}$=$\tau_{22}^{\eta}$						\\
 $\Lambda^{\bf8}_{5/2}$	& $m_2$		& $D_{05}^{\pi \Sigma}$=$\frac19 \left(2 \tau_{22}^{\pi} + 7 \tau_{32}^{\pi}\right)$	\\
			& 		& $D_{05}^{\eta \Lambda}$=$\tau_{22}^{\eta}$						\\
 $\Lambda^{\bf1}_{1/2}$	& 		& $S_{01}^{\bar K N}$=$\tau_{\frac120}^{\bar K}$		\\
 $\Lambda^{\bf1}_{3/2}$	& 		& $D_{03}^{\bar K N}$=$\tau_{\frac322}^{\bar K}$		\\
 $\Lambda^{\bf1}_{5/2}$	& 		& $D_{05}^{\bar K N}$=$\tau_{\frac522}^{\bar K}$		\\
													\\
													\\
\vspace{4.5pt}\\
\hline\hline
 \end{tabular}
 \quad
\begin{tabular}{ccl}
\hline\hline
 State		& Pole mass	& Partial wave, $K$-amplitudes					\\
 \hline
 $\Sigma^{\bf8}_{1/2}$	& $m_0$, $m_1$	& $S_{11}^{\pi \Lambda}$=$\frac13\tau_{10}^{\pi}$		\\
			& 		& $S_{11}^{\pi \Sigma}$=$\frac23\tau_{10}^{\pi}$		\\
			& 		& $S_{11}^{\eta \Sigma}$=$\tau_{00}^{\eta}$		\\
 $\Sigma^{\bf8}_{3/2}$	& $m_1$, $m_2$	& $D_{13}^{\pi \Lambda}$=$\frac16 \left(\tau_{12}^{\pi}+\tau_{22}^{\pi}\right)$		\\
			& 		& $D_{13}^{\pi \Sigma}$=$\frac13 \left(\tau_{12}^{\pi}+\tau_{22}^{\pi}\right)$		\\
			& 		& $D_{13}^{\eta \Sigma}$=$\tau_{22}^{\eta}$						\\
 $\Sigma^{\bf8}_{5/2}$	& $m_2$		& $D_{15}^{\pi \Lambda}$=$\frac{1}{27} \left(2 \tau_{22}^{\pi} + 7 \tau_{32}^{\pi}\right)$		\\
			& 		& $D_{15}^{\pi \Sigma}$=$\frac{2}{27} \left(2 \tau_{22}^{\pi} + 7 \tau_{32}^{\pi}\right)$		\\
			& 		& $D_{15}^{\eta \Sigma}$=$\tau_{22}^{\eta}$						\\
\hline
 $\Xi^{\bf8}_{1/2}$	& $m_0$, $m_1$	& $S_{11}^{\pi \Xi}$=$\frac19\tau_{10}^{\pi}$		\\
			& 		& $S_{11}^{\eta \Xi}$=$\tau_{00}^{\eta}$		\\
 $\Xi^{\bf8}_{3/2}$	& $m_1$, $m_2$	& $D_{13}^{\pi \Xi}$=$\frac{1}{18}\left(\tau_{12}^{\pi}+\tau_{22}^{\pi}\right)$	\\
			& 		& $D_{13}^{\eta \Xi}$=$\tau_{22}^{\eta}$					\\
 $\Xi^{\bf8}_{5/2}$	& $m_2$		& $D_{15}^{\pi \Xi}$=$\frac{1}{81}\left(2 \tau_{22}^{\pi} + 7 \tau_{32}^{\pi}\right)$	\\
			& 		& $D_{15}^{\eta \Xi}$=$\tau_{22}^{\eta}$					\\
 $\Xi^{\bf1}_{1/2}$	& 		& $S_{11}^{\bar K \Sigma}$=$\frac14\tau_{\frac120}^{\bar K}$		\\
			& 		& $S_{11}^{\bar K \Lambda}$=$\frac34\tau_{\frac120}^{\bar K}$		\\
 $\Xi^{\bf1}_{3/2}$	& 		& $D_{13}^{\bar K \Sigma}$=$\frac14\tau_{\frac322}^{\bar K}$		\\
			& 		& $D_{13}^{\bar K \Lambda}$=$\frac34\tau_{\frac322}^{\bar K}$		\\
 $\Xi^{\bf1}_{5/2}$	& 		& $D_{15}^{\bar K \Sigma}$=$\frac14\tau_{\frac522}^{\bar K}$		\\
			& 		& $D_{15}^{\bar K \Lambda}$=$\frac34\tau_{\frac522}^{\bar K}$		\\
\hline\hline
\end{tabular}\caption{Large $N_c$ mass eigenvalues in the $1/N_c$ expansion corresponding to states in the  $[{\bf ``20"},1^+]$ multiplet and the partial-wave amplitudes containing resonances with the same quantum numbers. Since $I=I'$ in all the partial-waves, to lighten notation we replaced the $II'Y$ labels in $\tau$ to those of a meson having the corresponding quantum numbers. Partial-wave amplitudes containing the nonstrange resonances were calculated before in Ref.~\cite{Cohen:2003nn} and are listed here for the reader's convenience. }
\label{TabAppB10}
\end{table}


\section{Reduced matrix elements for mixed-symmetric cores}\label{AppD}
In this Appendix we present the expressions for the reduced matrix elements of the $SU(6)$ generators for the mixed-symmetric cores. We built the MS cores as a inner core of $N_c-2$ quarks in a symmetric configuration coupled to an excited quark. The core wave function is a linear combination of states with definite inner core spin and it can be written as expression on Eq.~(\ref{coreswf}). Given this expression, the matrix elements for the $S_c$ operator can be expressed as
\begin{eqnarray}
 \langle S_c, {\bf R}_c ||S_c|| S'_c, {\bf R}'_c \rangle &=& \sum_{i,j} d_i d_j \delta_{R_{\tilde c_j},{\bf R}_{\tilde c_i}} \delta_{{\bf R}_c,{\bf R}'_c} \delta_{S_{\tilde c_j},S_{\tilde c_i}} (-1)^{S'_c+S_{\tilde c_i}+3/2}\hat S_c \hat S'_c\sqrt{D({\bf R}_c)}\nonumber\\
&& \times
 \left\{
    \begin{array}{ccc}
        S_c		& S'_c			& 1 \\
        S_{\tilde c_j}	& S_{\tilde c_i}	& 1/2
    \end{array}
\right\} \sqrt{S_{\tilde c_i}(S_{\tilde c_i}+1)(2S_{\tilde c_i}+1)}.
\end{eqnarray}

In the case of the $T_c$ operator, the reduced matrix elements for mixed-symmetric cores is given by
\begin{eqnarray}
 \langle S_c,{\bf R}_c ||T_c|| S'_c,{\bf R}'_c \rangle &=& \sum_{i,j} d_i d_j \frac{\hat S_c}{\sqrt{D({\bf R}_c)}} \delta_{S_c S'_c} \delta_{S_{\tilde c_i} S'_{\tilde c_j}} \delta_{{\bf R}_{\tilde c_i} {\bf R}'_{\tilde c_j}}\\
 &\times & \left[
(-1)^{\delta _{(q_{\tilde c_i}0)}} \sqrt{C_{SU(3)}({\bf R}_{\tilde c_i})}
 \left\{
    \begin{array}{ccc}
        {\bf R}_{\tilde c_i}	& {\bf 8}	& {\bf R}_{\tilde c_i} \\
        {\bf 3}		& {\bf 1}	& {\bf 3} \\
        {\bf R}'_c		& {\bf 8}	& {\bf R}_c
    \end{array}
\right\}
+
\frac{2\sqrt{3}}{3}
\left\{
    \begin{array}{ccc}
        {\bf R}_{\tilde c_i}	& {\bf 1}	& {\bf R}_{\tilde c_i} \\
        {\bf 3}		& {\bf 8}	& {\bf 3} \\
        {\bf R}'_c		& {\bf 8}	& {\bf R}_c
    \end{array}
\right\}
\right],\nonumber
\end{eqnarray}
where ${\bf R}_{\tilde c_i}=(p_{\tilde c_i},q_{\tilde c_i})$
and $C_{SU(3)}({\bf R})$  is the quadratic Casimir operator for $SU(3)$ given by $C_{SU(3)}(p,q)=\frac{p^2+q^2+pq+3p+3q}{3}$.

Reduced matrix elements for operator $G_c$ for mixed-symmetric cores can be expressed as
\begin{eqnarray}
&& \langle S_{c_{i}},{\bf R}_{c_{i}} || G_c || S_{c_{i'}},{\bf R}_{c_{i'}} \rangle_{\gamma_a}=
\sum_{j,j'}d_j d_{j'}\frac{\sqrt{6} \hat S_{c_{i}}\hat S_{c_{i'}}}{\sqrt{D({\bf R}_{c_{i}})}\sqrt{D({\bf R}_{\tilde c_{j}})}}\nonumber\\
&&\times \left[
- \delta_{S_{\tilde c_{j}}S_{\tilde c_{j'}}} \delta_{{\bf R}_{\tilde c_{j}}{\bf R}_{\tilde c_{j'}}}\sqrt{D({\bf R}_{\tilde c_{j}})}\hat S_{\tilde c_{j}}
\left\{
    \begin{array}{ccc}
        S_{\tilde c_{j}}	& S_{\tilde c_{j'}} 	& 0   \\
        1/2 			& 1/2			& 1   \\
        S_{c_i} 		& S_{c_{i'}}		& 1
    \end{array}
\right\}
\left\{
    \begin{array}{ccc}
        {\bf R}_{\tilde c_{j'}}	& {\bf 1}	& {\bf R}_{\tilde c_{j}}   \\
        {\bf 3}		& {\bf 8}	& {\bf 3}   \\
        {\bf R}_{c_{i'}}	& {\bf 8}	& {\bf R}_{c_i}
    \end{array}
\right\}\right.\nonumber\\
&& \left. +
\left\{
    \begin{array}{ccc}
        S_{\tilde c_{j}}	& S_{\tilde c_{j'}} 	& 1   \\
        1/2 			& 1/2			& 0   \\
        S_{c_i} 		& S_{c_{i'}}		& 1
    \end{array}
\right\}
\sum_{\gamma_b}
\left\{
    \begin{array}{ccc}
        {\bf R}_{\tilde c_{j'}}	& {\bf 8}	& {\bf R}_{\tilde c_{j}}   \\
        {\bf 3}			& {\bf 1}	& {\bf 3}   \\
        {\bf R}_{c_{i'}}	& {\bf 8}	& {\bf R}_{c_i}
    \end{array}
\right\}
\langle S_{\tilde c_{j}},{\bf R}_{\tilde c_{j}} || G_{\tilde c} || S_{\tilde c_{j'}},{\bf R}_{\tilde c_{j'}} \rangle_{\gamma_b}
\right]
\end{eqnarray}

The matrix elements of the $G_{\tilde c}$ operator  $\langle {\bf R}_{\tilde c_{j}}, S_{\tilde c_{j}} || G_{\tilde c} || {\bf R}_{\tilde c_{j'}}, S_{\tilde c_{j'}} \rangle_{\gamma_b}$ are given by expressions in Eq.~(\ref{AppB5}) and Eq.~(\ref{AppB6}) with replacement $N_c\to N_c-1$.

\end{document}